\documentclass[a4paper, 12pt]{article}
\usepackage[polish, english]{babel}
\usepackage[longnamesfirst]{natbib}
\usepackage[dvipsnames]{xcolor}
\usepackage{amsmath}
\usepackage{titling}
\usepackage{amsthm}
\usepackage{pdflscape}
\usepackage{amsfonts}
\usepackage[margin=1in]{geometry}
\usepackage{setspace}
\usepackage{enumitem}
\usepackage{booktabs}
\usepackage{caption}
\usepackage{multirow}
\usepackage{array}
\newcolumntype{C}[1]{>{\centering\arraybackslash}p{#1}}
\usepackage{float}
\usepackage{etoolbox}
\usepackage{sgame}
\usepackage{graphicx}
\usepackage{subfig}
\usepackage[retainorgcmds]{IEEEtrantools}
\usepackage{pgf}
\newcommand{\deriv}[1]{\frac{\partial}{\partial #1}}
\newcommand{\tderiv}[1]{\frac{\mathrm{d}}{\mathrm{d} #1}}
\usepackage{etoolbox}
\newcounter{deferred}
\newcommand{\deferred}[2][]{%
  \ifstrempty{#1}
    {\stepcounter{deferred}\expandafter\gdef\csname temp\arabic{deferred}\endcsname{#2}}
    {\expandafter\gdef\csname temp#1\endcsname{#2}}%
}
\newcommand{\shownow}[1]{\csname temp#1\endcsname}
\theoremstyle{definition}
\newtheorem{defi}{Definition}

\newtheorem{prop}{Proposition}

\theoremstyle{remark}

\DeclareMathOperator*{\argmax}{arg\,max}
\author{Pawe\l{} Gola\thanks{University of Edinburgh, Edinburgh, EH8 9JT, United Kingdom. E-mail: pawel.gola@ed.ac.uk}{ }  and Yuejun Zhao\thanks{University of Edinburgh and the Ragnar Frisch Centre for Economic Research. E-mail: yuejun.zhao@ed.ac.uk}}
\title{A Firm Link: \\ Overall, Between- and Within-Firm Inequality \\ Through the  Lens of a Sorting Model\thanks{We would like to thank Vasco Carvalho for inspiration and early discussions, as well as Jesper Bagger, Diego Battiston, Philipp Kircher,  Espen R. Moen, Isaac Sorkin and the seminar audience in Edinburgh for helpful comments and suggestions. Gola gratefully acknowledges the support from Iceland, Liechtenstein, and Norway through the European Economic Area (EEA) Financial Mechanism 2014–2021 (Project “Micro-level responses to socio-economic challenges in face of global uncertainties”, No S-BMT-21-8 (LT08-2-LMT-K-01-073)) under a grant agreement with the Research Council of Lithuania. Zhao acknowledges financial support from the Research Council of Norway, grant number 300917. Declarations of interest: none. All errors are ours.}} %
\date{December 2023}
\usepackage[pdftex]{hyperref}
\hypersetup{colorlinks=true, urlcolor=blue, citecolor=blue, linkcolor=black, pdfauthor={Pawel Gola}, final}
\usepackage[all]{hypcap}

\usepackage{xr}

\begin{document}
\setlength\emergencystretch{\hsize}
\maketitle
\begin{abstract}
This paper provides a new theory of the observed co-movement between overall wage inequality and its between-firm component. We develop and solve analytically a frictionless sorting model with two-sided heterogeneity, in which firms consist of distributions of tasks, choose how many workers to employ and reward their workers both through wages and amenities. We show that, for empirically-relevant parameter ranges, overall and between-firm inequality are firmly linked: A change in any of the models' primitives increases overall wage inequality if and only if it also increases the ratio of between-firm to overall inequality. Subsequently, we calibrate the model to match the Norwegian economy and find that the increase in wage inequality from 1995 to 2014 had a different primary cause (raising span-of-control cost) than the accompanying rise in welfare inequality (increased skill variance), and that the apparent decrease in wage inequality after 2015 masked a continued increase in welfare inequality.

\smallskip


\smallskip

\end{abstract}
\newpage

\section{Introduction}
\shortcites{Tomaskovic-Devey2020-dp}
\shortcites{Criscuolo2020-kr}
\shortcites{Hoen2022-sf}

Most developed countries experienced a well-documented increase in wage inequality from the 1960s until early 2010s. More recently,   \citet{Barth2016-lt,Barth2018-vq,Song2018,Sorkin2023-kb} have documented that in the United States, the increase in overall wage inequality has been driven mostly by a rise in between-firm wage inequality, with a much smaller contribution from changes in within-firm wage inequality; the same trend has been subsequently found to hold in a wide range of developed countries \citep{Tomaskovic-Devey2020-dp}.

This empirical trend is quite intriguing. First, its drivers are unclear. The sustained increase in overall inequality has previously been convincingly linked to skill-biased technological change, that is, to changes in technology that increase the return to skill \citep[see, e.g.,][]{Acemoglu2002-mn}. However, existing theories are silent on why skill-biased technological change would increase between-firm inequality by a larger proportion than within-firm inequality.\footnote{\cite{Cortes2023-zv} point out that skill-biased technological change contributes to the observed increase in between-firm wage inequality in a \cite{Melitz2003} model with search frictions, but even in this case it is unclear why this increase would be disproportionately large compared to the increase in within-firm inequality.} Second, paired with the recent findings by \cite{Sorkin2018} that as much as 70\% of firms' contribution to wage variance is caused by compensating differentials, the fact that most of the increase in overall wage inequality has been caused by its between-firm component creates the possibility that a significant part of the observed large increases in wage inequality may reflect changes in the compensating differentials paid by firms, and have no bearing on welfare inequality.%

In this paper, we attempt to determine how welfare inequality has evolved in the past three decades, and what are the main drivers of the observed changes in overall, between-firm and within-firm wage inequality. We take a structural approach to this problem, by first developing a novel, extremely tractable model of worker's sorting between- and within-firms, and then calibrating it to Norwegian administrative data.
Our model builds on \cite{Costrell2004} and \cite{Eeckhout2018}. As in \cite{Costrell2004}, firms consist of a hierarchy of tasks, that need to be performed in fixed proportions in order to produce output. As more complex tasks are complements with highly skilled workers, this implies that each firm hires a distribution of workers, and solves a within-firm problem of assigning workers to tasks. Further, we follow \cite{Eeckhout2018} in (a) allowing firms to be heterogeneous in their overall productivity and (b) choosing their size endogenously, subject to a span-of-control cost.  Finally, workers care not just about wages, but also the effort they exert, and the level of amenities provided by the firm, and effort and amenities enter the firm's profit function differently than wages.

Of course, a general model of heterogeneous, hierarchical firms with endogenous amenity provision would be entirely intractable. To ensure tractability, we impose three sets of simplifying assumptions. First,
we assume that the type of the firm and the type of the task enter the production function through a one-dimensional index, which we call the \emph{job} characteristics.\footnote{Think of a job as a firm-task pair; our simplifying assumption is then to impose a total order on these firm-task pairs.} Second, we impose a set of standard functional form assumptions on the production and utility functions which ensure that, for a given distribution of jobs and the optimal solution of the effort exertion problem, the sorting problem reduces to the problem of sorting workers to jobs---that is, it effectively reduces to a standard \cite{Sattinger1979} problem. However, the economy-wide distribution of jobs is a mixture of the within-firm job distributions weighted by the product of the number of firms of a given type and the firms' endogenous size, and thus remains an equilibrium object which can be found only by solving a complicated integral equation. Hence, and third, we assume that (a) workers' skill, (b) tasks within each firm, and (c) firms' productivity are all normally distributed with mean zero, which allows us to leverage the fact that a normal mixture of normal distributions is normal. Together with a standard Cobb-Douglas span-of-control cost function, the normality assumption ensures that the equilibrium distribution of jobs is always normal with mean zero; hence, one needs only to solve for the equilibrium variance of jobs, which is a problem that admits an analytical solution.

The equilibrium of the model reproduces a number of prominent empirical patterns. First,  even though the model is perfectly competitive, larger firm pay higher wage premia in equilibrium \citep[see, e.g.,][]{Bloom2018}. Notably, while all workers receive the same utility in all firms, the composition of their compensation differs systematically with the firm type. In particular, higher amenity provision decreases the dollar-cost of effort exertion, which raises optimal effort exertion more than proportionately. Hence, firms that provide higher amenities must additionally compensate their workers for the effort by paying them more. Because effort and firm productivity act as complements, more productive firms benefit more from cheaper effort exertion, and hence find it optimal to provide more amenities. At the same time, more productive firms find it optimal to hire more workers, so that in equilibrium larger firms pay workers higher wage premia and provide more amenities, both as a compensation for higher effort exertion.  Second, every type of firm hires all types of workers, but the workers hired by more productive firms are more skilled on average. Thus, the correlation between the skill of workers and the wage premia paid by firms is (a) equal to the correlation between workers skill and firm productivity, (b) strictly positive and (c) strictly smaller than 1. This means that the model is able to reproduce the findings from the literature on worker- and firm-fixed effect \citep[started by ][]{Abowd1999}, which---once the limited mobility bias is correctly accounted for---consistently finds positive, but small correlations between worker and firm fixed effects \citep[see ][for an in-depth discussion]{Bonhomme2023}.

As the model admits an analytical solution, we are able to derive comprehensive comparative statics results. We consider four exogenous changes to the primitives: (i) changes  in the distribution of workers' skill, (ii) changes in the distribution of firms' productivity, (iii) changes in the cost of amenity  provision and (iv) changes in the span-of-control cost. The main insight emerging from these results is that---for empirically relevant parameter ranges---each of these exogenous changes affects overall, within- and between-firm wage and welfare inequalities \emph{in the same direction}, but its effect on between-firm inequality is disproportionately large compared to its impact on within-firm inequality. In this sense, the empirical co-movement of wage inequality and its share explained by the between-firm component should not be surprising; indeed, in our model, increases in wage inequality are \emph{not} driven by the between-firm component only if the economy is affected by two or more exogenous changes that offset each other.

To understand the intuition behind this strong result, suppose that amenity provision is prohibitively expensive, and thus firms pay no wage premia. In that case, a worker's wage is an increasing function of their skill only, and---because more productive firms hire better workers on average---the average wage paid by a firm is an increasing function of the firm's type. Naturally, then, one can always \emph{re-normalise} the skill and productivity distribution in such a way that wages are linear in skill and average firm-level wages are linear in productivity.\footnote{In our case this is not necessary, as wages are linear in (squared) skill, and average firm-level wages are linear in (squared) productivity.} The following relationship among between-firm wage inequality, overall wage inequality and equilibrium sorting follows then easily from basic properties of covariance and the definition of correlation:
\begin{equation}
\frac{\text{Between-firm Wage Inequality}}{\text{Overall Wage Inequality}}=\text{Corr}(\text{Skill}, \text{Productivity})^2.
\end{equation}
Thus, the share of overall wage inequality that is explained by between-firm wage inequality is equal to the square of the equilibrium correlation between workers' skill and firms' productivity, both appropriately defined.\footnote{That is, defined in such a way that, in equilibrium, wages are linear in skill, and average firm-level wage is linear in productivity.}

In other words, the extent to which between-firm inequality accounts for overall inequality is very tightly linked to the strength of sorting between high skilled workers and high productivity firms. In our model, the strength of sorting is a function of  the equilibrium supply of quality jobs only.  The intuition for this is subtle, and thus we defer a detailed explanation until Section \ref{sec: sorting}; broadly speaking, however, the reason is that higher productivity firms consist of a better distribution of jobs than lower productivity firms, and thus both the strength of sorting and the supply of quality jobs are determined by how many of the jobs available in the economy are offered by highly productive firms. Because high skill workers and quality jobs are complements, any improvement in the supply of quality jobs increases overall wage inequality. Therefore, if a change in one of the model's primitives increases the ratio of between-firm to overall wage inequality, then it must increase the strength of sorting, which is only possible if it improves the supply of quality jobs; thus, it must also increase overall wage inequality.

The presence of compensating differentials complicates the relationship between overall and between-firm wage inequality, because the wage earned by a worker depends now also on the firms' type. This implies, in particular, that the ratio of between-firm to overall wage inequality depends not only on the strength of sorting, but also on the sign and size of the compensating differential paid by more productive firms. Nevertheless,  it remains true for empirically relevant parameter values that if a change in any of the four primitives considered by us increases wage inequality, it must also increase the ratio of between-firm to overall wage inequality.\footnote{Specifically, this result holds if (log) within-firm wage inequality is less than approximately $0.7$, which is true for instance, in all countries studied by \cite{Tomaskovic-Devey2020-dp}.}
Overall, therefore, our theoretical analysis reveals that the striking empirical co-movement of overall and between-firm wage inequality is the most natural outcome of the model, and can be caused by a wide range of exogenous changes. To shed some light on what its actual causes could have been, we need to bring the model to the data.

\begin{figure}
    \centering
    \includegraphics[width=\textwidth]{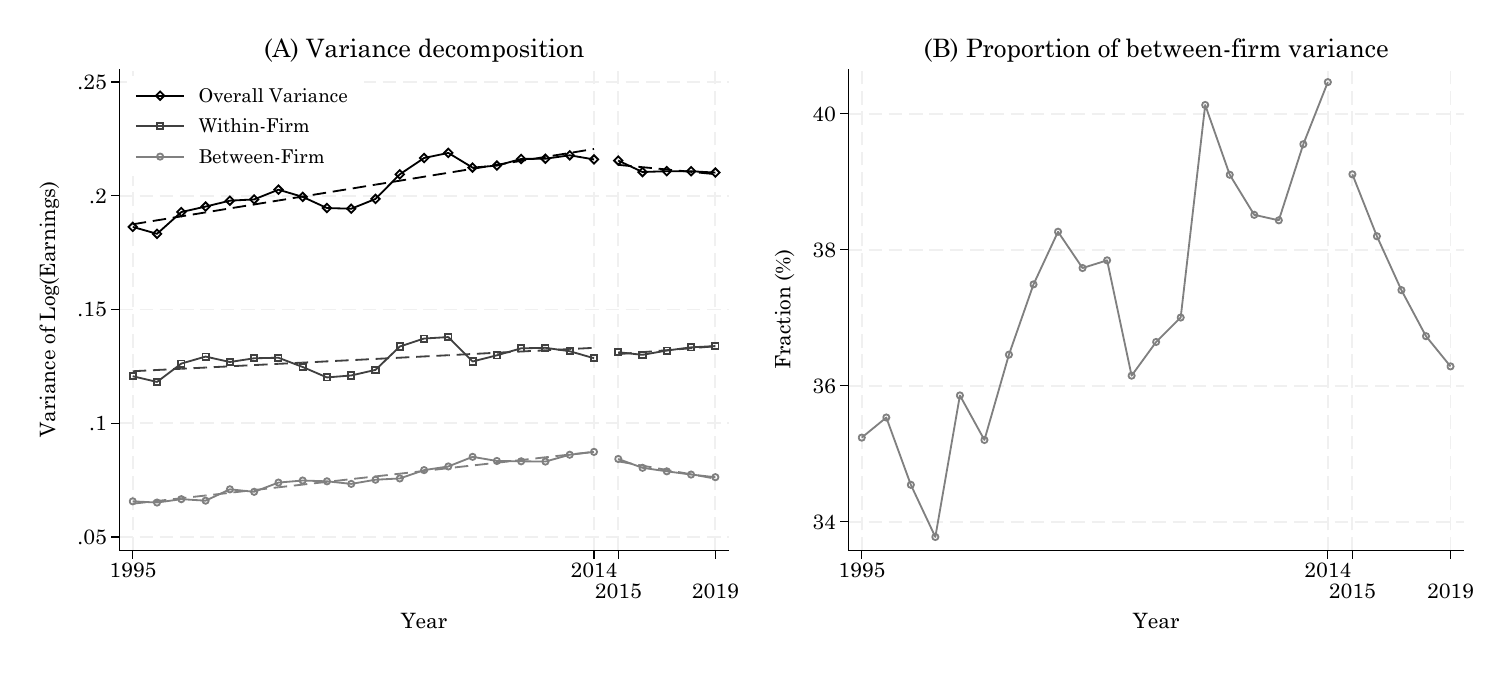}
    \caption{Decomposition of the variance of log annual earnings within and between firms in Norway from 1995 to 2019. \textit{(A)} Overall variance, within-firm variance, and between-firm variance. \textit{(B)} Proportion of between-firm variance.}
    \label{f:var_decomp}
\end{figure}

Specifically, we use our model to analyse the causes of changes in wage inequality between 1995 and 2019 in Norway. We start by documenting the empirical trends in the variance of wages, as well as within- and between-firm wage inequality in Norway in Figure \ref{f:var_decomp}. Between 1995 and 2014, which is the part of our panel that overlaps with the period analysed in \cite{Song2018}, the trends were qualitatively similar to those in the United States.\footnote{Of course, the levels are much lower in Norway than the US, but comparable to other Scandinavian countries such as Denmark \citep{Friedrich2022-sw} and Sweden \citep{Bonhomme2019-lg}.} Specifically, we find that overall wage variance increased by 3 log points,\footnote{We borrow the term `log points' from \citet{Song2018}. It means that the change in the variance of log earnings is multiplied by 100.} within-firm wage inequality increased by 0.8 log points, between-firm wage inequality increased by 2.2 log points, implying that the share of overall inequality explained by  between-firm wage inequality increased by 5 percentage points. %
 The results for years 2015 to 2019, which we are first to document, show a reversal in some of these trends.\footnote{Due to a change in the reporting scheme in 2015 \citep{ssb2014-change}, we divide our sample into two parts, 1995 to 2014, and 2015 to 2019. See Section \ref{ssec: data} for more details.} While within-firm wage inequality has continued with a modest increase of 0.3 log points, overall wage inequality experienced a decrease of 0.5 log points, and between-firm wage inequality has fallen by 0.8 log points, leading to a sharp decrease in the share of overall inequality explained by  between-firm wage inequality by 3 percentage points.

We calibrate the model to matched employer-employee data in Norway. Our model can be identified using four key moments: within-firm wage inequality, between-firm wage inequality, the variance of within-firm wage variances and the average within-firm variance of wages \emph{unweighted} by firm size.\footnote{The difference between the first and fourth moment is that in the first moment is calculated by taking a weighted average of the within-firm variance of wages.}  We find that from 1995 to 2014, the variance of skill has sharply increased, whereas the span-of-control cost has sharply decreased, with little to no change in the distribution of firm productivity and the cost of amenity provision. Taken together, these changes have improved the  equilibrium supply of quality jobs. After 2015, in contrast, the supply of quality jobs has decreased in the Norwegian economy, which was driven by a fall in the variance of productivity and an increase in the cost of amenity provision.

With a calibrated model in tow, we simulate changes in welfare inequality over the entire period and find that the trends in welfare inequality qualitatively match trends in wage inequality from 1995 to 2014. However, after 2015 overall welfare inequality continues with a modest increase. Hence, our calibration indicates that post-2014 the observed decreases in overall wage inequality reflect shifts in the composition of compensation between wages and compensating differentials, rather than actual decreases in underlying welfare inequality.

Finally, we perform counterfactual simulations to isolate the impact that each of the four channels had on wage and welfare inequality over the period of analysis. For 1995-2014 we find that  while a fall in the span-of-control cost was the main driver of the increase in wage inequality (62\%), changes in the distribution of workers skill were the main contributor to the increase in welfare inequality (52\%).
Given that changes in the distribution of skill are isomorphic to skill-biased technological change in our model, this finding reconciles the old and new literatures on inequality: Changes in the price/variability of skill were indeed the main driver of changes in \emph{welfare} inequality, because a large proportion of the increase in between-firm wage inequality was welfare-neutral---specifically, out of the 62\% of the increase in overall wage inequality that was caused by a fall in the span-of-control cost, more than three quarters reflects changes in the compensating differential for effort exertion. %

Post-2014, the emerging deterioration in the quality of jobs in the Norwegian economy has dominated the continued increase in the variance of skill, leading to a decrease in overall wage inequality. However, for realistic parameter values, changes in the distribution of skill have a stronger effect on welfare than on wage inequality; thus, the increase in the variability of skill has caused a continued increase in welfare inequality post-2014.

The rest of the paper is structured as follows. Section \ref{sec: rellit} discusses the related literature. Section~\ref{sec: model} develops the model and discusses our modelling choices. Section \ref{sec: eqmchar} solves the model. Section \ref{sec: sortingwagesutility} discusses the behaviour that emerges in equilibrium and provides comparative statics results. Section \ref{sec: calibration} calibrates the model to the Norwegian data, simulates the evolution of welfare inequality, and provides a counterfactual analysis. Section \ref{sec: conrem} concludes. Appendix \ref{sec: omm} contains omitted proofs and derivations. Online Appendix \ref{sec: WFEFFE} discusses how to conduct an AKM variance decomposition in the context of our model. Online Appendix \ref{app: bootstrap} details the resampling procedure used to calculate confidence intervals.   Online Appendix \ref{sec: addfigures} contains additional figures relating to the calibration exercise.

\section{Related Literature}\label{sec: rellit}

\paragraph{Theoretical Sorting Models}

By merging the models of \cite{Costrell2004} and \cite{Eeckhout2018} our paper contributes to the literature on labour market sorting: It introduces  firm heterogeneity and firm's size choice into the former and hierarchical firms (i.e., firms that solve a within-firm assignment problem) into the latter.  \cite{Eeckhout2018} itself nests a large family of sorting models in which the size of a firm/number of workers performing a task is endogenous, as in \cite{Sattinger1975, Teulings1995, Teulings2005, Costinot2009, Costinot2010}. Our paper provides a blueprint on how to introduce within-firm heterogeneity into these models, while maintaining tractability. Indeed, even with the addition of the within-firm sorting problem, our choice of normal skill and productivity distributions, coupled with a production function that makes equilibrium firm size exponential in the square of firm's type yields a more tractable model than most in this literature.\footnote{A notable exception is \cite{Teulings1995}, whose model also allows for a fully analytical solution, and from whose functional form choices we draw considerable inspiration.}

This paper contributes to the nascent literature that uses sorting and/or search models to investigate the causes of the more-than-proportional increase in between-wage inequality. \cite{Boerma2023} and \cite{Freund2022} both develop models of team formation in which the number of team members is fixed; hence, these models do not feature the connection before firms' size and equilibrium sorting, which is critical for the tight link between overall and between-firm wage inequality highlighted in this paper. \cite{Trottner2022-hv} and \citet{Cortes2023-zv} study, respectively, the impacts of trade liberalisation and skill-biased technological change on overall, between- and within-firm inequality, in models that build on \cite{Melitz2003}. In both of these models, firms' production is determined endogenously and better firms tend to be both larger and hire better workers; both papers emphasise a similar mechanism, by noting that their shock of interest increases the employment in the most productive firms and through that raises between-firm inequality.\footnote{\cite{Eeckhout2014} also develop a model in which all firms hire workers of many types, and more productive firms are larger and hire better workers. However, \cite{Eeckhout2014} do not apply their model to the study of wage inequality. }
 However, by developing a model with a simple analytical solution and considering a range of possible changes of primitives, we are able to make a more general point: The relationship between the size of productive firms and sorting means that most shocks (rather than just skill-biased technological change or trade liberalisation) which increase overall wage inequality do so by increasing between-firm inequality \emph{by a larger proportion} than within-firm inequality. Our flexible framework allows us also to quantify the extent to which different types of changes have affected overall, within and between-firm inequality.

\paragraph{Empirical Literature on Between-Firm Inequality}
Our paper contributes to the large literature documenting that the increase in overall wage inequality from the 1980s to early 2010s was disproportionately driven by the increase in its between-firm component.
The rising influence of between-firm wage inequality is documented for 12 (out of 14) developed countries (including Norway and the US) in \citet{Tomaskovic-Devey2020-dp}; %
 for the US in \citet{Barth2016-lt,Barth2018-vq,Song2018,Sorkin2023-kb}, and (for more recent years) \citet{Haltiwanger2022-zx,Haltiwanger2023-km};
for Sweden in \citet{Hakanson2021-fw};
 for Germany in \citet{Card2013,Baumgarten2020-rv, Freund2022, Lochner2022-iz};
and for Brazil (from 1986--1995)  in \citet{Helpman2016}.

Our paper contributes also to the growing body of work documenting that overall and between-firm wage inequality have both decreased in the second half of the 2010s in countries that have previously experienced increases in inequality. Specifically, overall wage inequality fell slightly between 2012 and 2020 in the US when measured using weekly earnings \citep{Aeppli2022-un},\footnote{With respect to hourly and annual earnings, \citet{Aeppli2022-un} find that inequality has stabilised since 2012 in the US.}
between 2014 and 2016 in Sweden \citep{Engbom2023-pd},
and between 2010 and 2017 in Germany \citep{Lochner2020-wf, Freund2022}.
Accordingly, between-firm wage inequality experienced a slight decline between 2014 and 2018 in the US \citep{Landefeld2023}, and in the late 2010s in Germany \citep{Freund2022}.

\paragraph{Amenities and Inequality}
In our model, the only reason for firm wage premia are the compensating differentials for elevated effort; this modelling choice is motivated by the finding in \citet{Sorkin2018} that compensating differentials account for majority of the variance in wage premia. Further,  \citet{Ouimet2023-qr} document that the vast majority of variation in non-wage benefits---health, retirement, and leave benefits---is explained by the between-firm component; this finding has motivated our choice of modelling amenities to be the same for all workers within a firm.  There is a number of recent papers focusing on how much inequality there exists in amenity provision, and how this amenity inequality affects measures of overall inequality \citep{Kristal2020-wf, Sockin2022, Bana2023-sb, Maestas2023}.\footnote{In addition, \cite{Bagger2021} emphasise a less direct connection between amenities and inequality. They show, using a calibrated directed search model, that the presence of amenities amplifies sharply the deadweight loss of labour income taxation. Thus, in the presence of amenities the social planner should adopt lower tax rates, leading to more wage inequality.} However, with the exception of \cite{Sockin2022}, these studies look at a limited subset of the amenities that matter in reality and that our model allows for. Indeed, the most comparable numbers are those provided by \cite{Sorkin2018} and \cite{Taber2020}, in which a revealed preference approach is adopted.  \cite{Sorkin2018} and \cite{Taber2020} find that `real' inequality (i.e., inequality after accounting for the impact of compensating differentials) is about 23\% lower and 125\% higher, respectively, than observed wage inequality. In our calibrated model, welfare inequality is about 30\% lower than observed wage inequality.\footnote{The fact that this number is lower in our calibrated model is expected, given that we interpret any variation in firm wage premia to be caused by compensating differentials. Indeed, if all of firm variation is excluded, then the decrease in inequality calculated by \citet{Sorkin2018} is 28.5\%, and thus very similar to the number produced by our calibration.}

\paragraph{Skill-Biased Changes}
There is a large literature linking changes in inequality to either skill-biased technological change, or increases in the heterogeneity of workers' skill.\footnote{In our model, skill-biased technological change is isomorphic to changes in the distribution of skill.} An increase in the demand for cognitive skills caused by technological change has been identified as the main culprit behind the increase in wage inequality from the 1960s to the 1980s \citep{Bound1992, Katz1992, Juhn1993}. More recently, \citet{Lindner2022-gg} have shown that skill-biased technological change increase the college premium by 6.1\% in Norway from 2002 to 2013. \citet{Acemoglu2022-dx} document that more than half of the change in wage structure in the US in the past four decades was due to automation replacing low-skilled jobs, whereas about 10\% of the change resulted from skill-biased technological change. Finally, much of the AKM literature \citep[e.g., ][]{Card2013, Song2018} finds sizeable increases in the variance of worker fixed effects, which can be most readily explained by increases in the variance of skill or the return to skill.\footnote{It is worth noting, however, that in our model changes in any other primitive could also increase the variance of worker fixed effects, albeit many of them only indirectly.}

\paragraph{Firm-Side Changes}
\citet{Autor2020-ny} document the rise of superstar firms,  and show that (a) it explains declining labour shares in the United States and (b) that technological change is an important driver of the increasing concentration in certain industries.
\citet{Mueller2017-vr} provides direct evidence that the increases in inequality in the UK were caused by the increase in employment by the largest firms in the economy, which is consistent with the theoretical mechanism we highlight. %
 \citet{Barth2023-hw} show that capitalised software investments by firms cause increases in  both within- and between-firm wage inequality.

\section{Model}\label{sec: model}
There are two populations: workers and firms. Workers differ in skill.  Following \cite{Costrell2004} we assume that each firm consists of a hierarchy of tasks. Following \cite{Eeckhout2018} we additionally assume that firms differ in productivity and choose endogenously how many workers to employ. Finally, on top of all that, we introduce amenities provision by the firms.

\paragraph{Workers}

There exists a unit measure of workers. Each worker is endowed with skill $x$, which is distributed normally with mean 0 and variance $\sigma_x$.  %
A worker's utility is a product of the wage $w$ they receive \emph{per unit} of the effort $e$ they provide and the number/quality of amenities $a$ provided by the firm they are employed by, with

$$u(w, a)=\frac{w}{e}a.$$  Workers are paid competitive wages that depend on their skill, exerted effort and the amenity level provided by the firm. Their reservation utility is equal to 0.

\paragraph{Firms}

There exists a unit measure of profit-maximising firms. Firms differ in productivity $\theta$, which  is distributed normally with mean 0 and variance $\sigma_{\theta}$. All firms consist of (the same) collection of tasks $t$. Each firm decides on the number and types of workers they hire, the assignment of workers to specific tasks, and the number/quality of amenities the firm provides.   A worker of skill $x$ hired by a firm of productivity $\theta$ to exert effort $e$ in performing task $t$ contributes
\[q(x, t, \theta, e)=A\sqrt{\exp({x(t+\theta)})e},\]
to total output, where $A>0$ is the total factor productivity (TFP).

Following \cite{Costrell2004} we assume that for the final good to be produced, tasks need to be filled in an exogenously fixed proportion. Specifically, we assume that $\Phi(h)$ of workers employed by firm $\theta$ perform a task with $t \leq h$, where $\Phi$ denotes the cdf of a standard normal distribution.
Firms face a span-of-control cost $$C_l(L)\equiv L^{1+c_l},$$ and an amenity provision cost $$C_a(a, L)=\frac{L}{c_a} \left( \frac{c_a a}{1+c_a}  \right)^{1+c_a},$$ where $c_l, c_a$ capture the convexity of their respective cost functions. The amenity provision cost, in particular,  has been chosen in such a way that if $c_a \to \infty$ then, in equilibrium, all firms choose $a=1$.

Denoting by  $m: \mathbb{R} \to \mathbb{R}$ the assignment function of tasks to workers (i.e.,  $m(t)$ denotes the skill of the worker assigned to perform task $t$) and by $e: \mathbb{R} \to \mathbb{R}_{\geq 0}$ the effort function (i.e., $e(x)$ denotes the effort exerted by worker of skill $x$),  the total output of firm $\theta$ is equal to the sum of the output's of all tasks and the two costs:
\[Q(L, m, \theta, e)=L\int_{-\infty}^{\infty} q(m(t), t, \theta, e(m(t))) \mathrm{d} \Phi(t)-C_L(L)-C_A(a, L). \] The firms' reservation profit is 0.

\paragraph{Profit Maximisation and the Demand for Skill}
The demand for skill  is determined by firms' hiring, assignment, effort exertion and amenity provision decisions. In order to hire a worker of skill $x$ each firm has to provide them with the competitive utility $u(x)$, where $u: \mathbb{R}\to \mathbb{R}$. The need to provide worker of skill $x$ with utility $u(x)$, implies that they have to be offered wage $w(x, a, e)=u(x)e/a$. Overall, firm $\theta$ earns profit $r(\theta)$, provides amenities $a(\theta)$, fills task $t$ with a worker of skill $m(t; \theta)$, requires workers of skill $x$ to exert effort $e(x, \theta)$ and hires $L(\theta)$ workers, where
\begin{IEEEeqnarray}{rCl}
r(\theta) &=& \max_{L, m, a, e} \, Q(L, m, \theta, e(m))-  L \int_{-\infty}^{\infty} \frac{u(m(t))e}{a} \,  \mathrm{d} \Phi(t)\label{eq: profit} \\
(L(\theta), m(\theta), a(\theta), e(\theta)) &\in& \argmax_{L, m, a, e} \,   Q(L, m, \theta, e(m))-  L \int_{-\infty}^{\infty} \frac{u(m(t))e}{a} \,  \mathrm{d} \Phi(t). \label{eq: profitmax}
\end{IEEEeqnarray}

We will define the demand for skill analogously to  \cite{Gola2021}. The \emph{demand for skill} $x$, $D(x)$, is equal to the measure of tasks filled by workers with skill of \emph{at least} x:
\begin{equation}
D(x)=\int_{-\infty}^{\infty} L^*(\theta) \text{Pr}(m^*(H; \theta) \geq x) \mathrm{d} \Phi(\theta). \label{eq: demand}
\end{equation}
 Of course, the supply of skill $x$, $S(x)$, is defined as the measure of workers with skill greater or equal to $x$, with $S(x)=1-\Phi(x/\sigma_x)$.
\paragraph{The Competitive Equilibrium with Full Employment}
In the competitive equilibrium, firms maximise their profits given equilibrium utilities, and markets clear. For notational simplicity, and without loss of generality, we will restrict attention to equilibria with full employment.\footnote{Recall that an unemployed worker receives zero utility. Hence, they would be happy to work for any finite wage, receive any finite amenity and exert any finite effort. Similarly, any firm---regardless of the amenities it provides---would be happy to employ a worker of even the lowest skill, if the wage is low enough and the effort exerted is sufficiently high, as the revenue produced by such a worker would exceed the additional wage, amenity and span-of-control costs incurred. Thus, all workers must be employed in any equilibrium.}

\begin{defi}[Equilibrium with Full Employment]\label{defi: eqm}
A competitive equilibrium with full employment is characterised by:\newline
(a) a \emph{task assignment function} $m^*: \mathbb{R}^2 \to \mathbb{R}$, a \emph{firm-size function}  $L^*: \mathbb{R} \to \mathbb{R}$, an \emph{effort function} $e^*: \mathbb{R}^2 \to \mathbb{R}$, and an \emph{amenities provision function} $a^*: \mathbb{R} \to \mathbb{R}$, each consistent with firms' profit maximisation (Equations \eqref{eq: profit} and \eqref{eq: profitmax}); \newline
(b) a \emph{utility function} $u: \mathbb{R}\to \mathbb{R}$, which clears the market, that is, equates the demand and supply for skill: $$D(x)\equiv\int_{-\infty}^{\infty} L^*(\theta) \text{Pr}(m^*(H; \theta) \geq x) \mathrm{d} \Phi(\theta)=1-\Phi(x/\sigma_x)\equiv S(x).$$
\end{defi}

The wage paid in equilibrium to worker $x$ who works for firm $\theta$  is pinned down by the utility function, the effort function, and the amenities provision function with
\begin{equation}\label{eq: eqmwage}
    w(x, \theta)=\frac{u(x)e^*(\theta)}{a^*(\theta)}.
\end{equation}
\subsection{Discussion}\label{sec: diss}

Let us briefly discuss a few of our modelling and analysis choices.

\subsubsection{The Role of Effort} Effort exertion plays two roles in the model. First, and most importantly, the introduction of effort---together with the exponential production function---ensures that the logarithm, rather than the level, of wages and utility becomes the natural unit of analysis. This makes it much easier to calibrate the model to the variance of log wages, which is the most commonly used empirical measure of inequality. Second, the interplay between effort and amenity provision ensures that larger firms (a) pay higher wages and (b) provide better firm-level amenities in equilibrium, which aligns the model with recent empirical findings \citep{Lamadon2022, Sockin2022}.

It is wort noting that (the inverse of) `effort' could be more generally interpreted as any amenity chosen at the task-firm-worker level, as opposed to `amenities', which are chosen at the firm level. We chose the label `effort' to succinctly differentiate between these two types of amenities.

\subsubsection{Alternative Parameterisations} The model can be parameterised in a number of equivalent ways. It is well-known that in assignment/sorting models one can normalise either the distributions of characteristics or the production function. This remains true in our model. While we have decided to allow for  changes in the variance of the  distributions of skills and productivity, we could have equivalently assumed that skills and productivity are standard normally distributed, and---instead---the production function is as follows:
\[\bar{q}(x, t, \theta, e)=\sqrt{\exp({\sigma_x x(t+\sigma_\theta \theta)})e}.\]
This isomorphism is the reason why we occasionally interpret changes in $\sigma_x$ as skill-biased technological change---as it is equivalent to an improvement in the productivity of skills.

Further, once the alternative production function $\bar{q}$ has been written down, it becomes obvious that there exist many equivalent formulations, for example
\[\bar{q}(x, t, \theta, e)=\sqrt{\exp({\sigma_x x(t+\sigma_\theta \theta)})e}=\sqrt{\exp({\gamma x(\beta t+(1-\beta) \theta)})e}=\sqrt{\exp({ x(\sigma_x t+\zeta \theta)})e},\]
where $\sigma_x\equiv\gamma \beta$, $\sigma_\theta \sigma_x\equiv\gamma (1-\beta)\equiv\zeta$.

The rationale for choosing our baseline parametrisation was two-fold. First, we decided to keep the production function constant and change the distributions, because that makes $\sigma_x$ easily interpretable as a change in the variability of skills. Second, we decided to normalise the distribution of tasks within firms to be standard normal, and allow for changes in the distributions of skill and productivity, because this was leading to by far the easiest algebraic formulations and cleanest intuition. Nevertheless, it remains true that some potential real-world changes---such as changes to the proportion of tasks needed to produce a good---would be best represented in the model through a combination of changes to $\sigma_x$ and $\sigma_\theta$.

\subsubsection{The Role of $\sigma_x, \sigma_\theta$}\label{sec: variance}

Finally, it is worth noting that while $\sigma_x$ and $\sigma_\theta$ technically capture the variability of skill and productivity, respectively, they $\emph{de facto}$ also capture the average level of skill and productivity in the economy. The reason is that sorting between tasks and skills will be positive and assortative in equilibrium, and hence workers with negative skill will work for firms (and in tasks) with negative productivity. As the production function is multiplicative, this implies that workers with highly negative and highly positive skills are equally productive. In other words, it is $x^2$ and $\theta^2$ that truly capture skill and productivity in this model; and, of course, the more variable $x$ is, the higher is the mean of $x^2$.

The fact that the variance and level of the distributions of skill and productivity are captured by one parameter each is not ideal, but it is critical for the remarkable tractability of this model.

\section{Solving the Model}\label{sec: eqmchar}

In this Section, we will solve for the equilibrium of the model.
Let us start by tackling the assignment of workers to \emph{jobs} first, where a job is the sum the task and firm types:
\[h=t+ \theta.\]
Note that the equilibrium distribution of jobs in the economy is given by
\begin{equation}\label{eq: jobdistr}
F(h)\equiv \int_{-\infty}^{\infty}  L^*(\theta)\Phi(h-\theta)) \mathrm{d} \Phi(\frac{\theta}{\sigma_\theta}).
\end{equation}
We will \emph{guess} that $F(h)=\Phi(\frac{h}{\sigma})$, where $\sigma> \max\{\sigma_x, 1\}$, that is, we will guess that there exists an equilibrium in which jobs are normally distributed with mean 0 and a high enough variance. In the following step, we will derive the wage functions holding in equilibrium given this guess. Finally, we will show that these wage functions lead to firm's size choices that indeed result in the jobs being normally distributed in the economy. In the process, we will also solve for the equilibrium variance of jobs $\sigma^2$.\footnote{This equilibrium is unique in the class of equilibria in which jobs are normally distributed.} Finally, note that for the same reasons as those discussed in Section \ref{sec: variance}, $\sigma$ describes not only the variability of the job distribution but also, effectively, the average quality of supplied jobs. For that reason, we will typically refer to $\sigma$ as the \emph{supply of quality jobs}.

\paragraph{Effort exertion, Assignment and Utility}

Define the function $\mu: \mathbb{R}^2 \to \mathbb{R}$, which determines which worker performs job $h$ at firm $\theta$. Then the  first order condition (the Euler-Lagrange condition) of the firm's problem with respect to the effort schedule $e$ yields:

$$e^*(\mu(h), \theta)=\frac{A^2\exp{({\mu(h)h})}a^2}{4(u(\mu(h, \theta))^2}.$$
Note that this implies that firms offering higher amenities require their workers to provide disproportionately higher effort. Hence, perhaps surprisingly, firms offering higher amenities will need to offer higher wages to workers. In other words, firms compensate workers for exerting higher effort both by paying them higher wages and by providing better amenities.

Substituting the expression for optimal effort exertion into the firm's problem yields:

\[r(\theta) = \max_{a, L} \max_{\mu}\  L\int_{-\infty}^{\infty} \frac{A^2a\exp{({\mu(h)h})}}{4u(\mu(h, \theta)} \, \mathrm{d} \Phi(h-\theta)) -C_l(L) -C_a(a,L).\]

We can now solve the problem of assigning workers to jobs. Technically this is a two-dimensional problem, because the production function depends on jobs and---through amenity provision---on the firm's type. However, one can easily see that the the first order (Euler-Lagrange) condition with respect to the assignment $\mu$ implies that:
\begin{equation}\label{eq: FOC} \frac{u'(\mu(h, \theta))}{u(\mu(h, \theta))}=h,\end{equation}
which does not depend on the firm's amenity provision. It must be the case, therefore, that the optimally chosen $\mu(\cdot; \theta)$ is independent of $\theta$; henceforth, we will suppress $\theta$ when referring to $\mu$. Clearly, by standard arguments \citep[see, e.g.,][]{Sattinger1979}, the second-order condition is satisfied only if $\mu(h)$ is increasing. Further, full employment implies that $ \int_{-\infty}^{\infty}  L^*(\theta) \mathrm{d} \Phi(\theta)=1$ in equilibrium, so that
\[D(x)=1-\Phi(\frac{(\mu^*)^{-1}(x)}{\sigma}).\]
For the market to clear we require $D(x)=S(x)=1-\Phi(x/\sigma_x)$, implying that $\mu(h)=\frac{\sigma_x}{\sigma}h$ in equilibrium. Hence, by Equation \eqref{eq: FOC}, we can pin down the derivative of log utility, with
$\frac{u'(\frac{h}{\sigma})}{u(\frac{h}{\sigma})}=h.$
Integrating from 0 to $x$ gives the difference between the log utilities enjoyed by workers of skill $x$:
\begin{equation}
\ln(u(x))=  \frac{\sigma}{\sigma_x} \frac{x^2}{2}+\ln(u(0)).
\label{eq: utility}
\end{equation}

\paragraph{Amenities and Firms' Size} Using the optimal assignment, the expression for equilibrium utility, and the fact that
\[ E_h \left(\left.\exp\left(\frac{\sigma_x h^2}{2 \sigma}\right) \right | \theta\right)=\frac{\exp\left(\frac{\theta^2}{2(\frac{\sigma}{\sigma_x}-1)}\right)}{\sqrt{1-\frac{\sigma_x}{\sigma}}},\]
firm's $\theta$ profit can be further rewritten as
\begin{IEEEeqnarray}{rCl}
    r(\theta)&=&\max_{L, a}L \left (\frac{aA^2}{4u(0)}\frac{\exp\left(\frac{\theta^2}{2(\frac{\sigma}{\sigma_x}-1)}\right)}{\sqrt{1-\frac{\sigma_x}{\sigma}}}-\frac{1}{c_a} \left( \frac{c_a a}{1+c_a}  \right)^{1+c_a}-L^{c_l}\right).
\end{IEEEeqnarray}
It is then straightforward to show that firm's $\theta$ optimal choice of amenities and size are

\begin{IEEEeqnarray}{rCl}
  a^*(\theta)&=&(\frac{1}{c_a}+1)\left(\frac{A^2}{4u(0)}\frac{\exp\left(\frac{\theta^2}{2(\frac{\sigma}{\sigma_x}-1)}\right)}{\sqrt{1-\frac{\sigma_x}{\sigma}}} \right)^{\frac{1}{c_a}}, \label{eq: amenities}
  \\L^*(\theta)&=&\frac{\exp{\left [\alpha  \left (\frac{\theta^2}{2(\frac{\sigma}{\sigma_x}-1)}+2\ln A-0.5\ln (1-\frac{\sigma_x}{\sigma})-\ln(4u(0)) \right)\right]}}{(1+c_l)^{\frac{1}{c_l}}}, \label{eq: size}
\end{IEEEeqnarray}
where $\alpha\equiv \frac{1+c_a}{c_lc_a}$. The fact that the size chosen by a firm depends exponentially on the square of the firm's productivity is of critical importance, as it implies that (by Equation~\eqref{eq: jobdistr}), the distribution of jobs will resemble a normal mixture of normal distributions, and thus will itself be normal.

\paragraph{Job Distribution Revisited}
Specifically, by Equation~\eqref{eq: jobdistr} the density of the equilibrium job distribution is given by
\begin{equation}\label{eq: densityorig} f(h)=\int_{-\infty}^{\infty}  \frac{L^*(\theta)}{\sigma_{\theta}}\phi(h-\theta) \phi(\frac{\theta}{\sigma_{\theta}}) \, \mathrm{d} \theta,  \end{equation}
where $\phi(\cdot)$ denotes the pdf of the standard normal distribution. After a few rearrangements, the density of the job distribution can be written as:

\begin{IEEEeqnarray}{rCl}
f(h)= \frac{L^*(0)}{\sqrt{1+\sigma^2_{\theta}\left(1-\frac{\alpha}{\sigma/\sigma_x -1} \right) }} \phi  \left (h \sqrt{\frac{1}{1+\frac{\sigma_{\theta}^2}{1-\frac{\alpha\sigma_{\theta}^2}{\frac{\sigma}{\sigma_x}-1}}}} \right ).\footnote{The details of the derivation can be found in Appendix \ref{app: mixture}.} \label{eq: density}
\end{IEEEeqnarray}

For our guess to be correct we need $f(h)=\frac{1}{\sigma}\phi(\frac{h}{\sigma})$ implying that:
\begin{IEEEeqnarray}{rcl} 1&=& \frac{L^*(0)}{\sqrt{1-\frac{\alpha \sigma_{\theta}^2}{\frac{\sigma}{\sigma_x} -1}}}, \label{eq: eqmmass} \\
\sigma^2&=&\frac{\sigma_{\theta}^2}{1-\frac{\alpha\sigma_{\theta}^2}{\frac{\sigma}{\sigma_x}-1}}+1 \label{eq: eqmsigma}.
\end{IEEEeqnarray}
Equation \eqref{eq: eqmsigma} can be written as a third-degree polynomial, which can be solved analytically, with a unique solution which meets the restrictions $\sigma>\max\{1, \sigma_x\}$; however, the solution is so complicated, as to be unhelpful. Instead, I will solve this equation for $\alpha$, which yields:
\begin{equation}\label{eq: inversesol}
\alpha=t^{-1}(\sigma; \sigma_x, \sigma_\theta) \equiv \left(\frac{\sigma}{\sigma_x}-1\right) (\frac{1}{\sigma_{\theta}^2}-\frac{1}{\sigma^2-1}).
\end{equation}
As $\sigma^2>1$ and $\alpha>0$, it follows by inspection that $\frac{\partial}{\partial \sigma}t^{-1}(\sigma)>0$; thus, an inverse of $t^{-1}(\cdot)$ is well-defined and also strictly increasing. I will denote this inverse as $t$; clearly, $t(\alpha; \sigma_x, \sigma_\theta)=\sigma$ and so gives us the standard deviation of the equilibrium distribution of jobs.

The last step is to find the level of utility that ensures that firms want to hire exactly a measure one of workers. Using Equations \eqref{eq: size} and \eqref{eq: eqmmass} it can be shown that
\begin{equation}
    \label{eq: u0} \ln (u(0))=2 \ln(A) -0.5\ln(1-\frac{\sigma_x}{\sigma})-\ln 4 -\frac{1}{\alpha}\left(\ln(\sigma_{\theta})-0.5 \ln(\sigma^2-1) \right)-\frac{c_a}{1+c_a}\ln(1+c_l),
\end{equation}
which closes the model. Overall, therefore, there exists a single equilibrium in which jobs are normally distributed.\footnote{It remains an open questions whether there exist equilibria in which jobs are not normally distributed, although we conjecture that there are no such equilibria.}

\section{Jobs, Sorting and Inequality}\label{sec: sortingwagesutility}
In this Section, we first characterise the determinants of the supply of quality jobs $\sigma$ (Section \ref{sec: qualityjobs}), the sorting between high skill workers and high productivity firms (Section \ref {sec: sorting}), as well as welfare and wage inequality (Sections \ref{sec: welfareinequality} to \ref{sec: diffimpact})  in our economy. %

\subsection{Supply of Quality Jobs}\label{sec: qualityjobs}
It is natural to start our analysis by investigating how the primitives of the model determine $\sigma$, as this is the main endogenous variable in our model.

The key observation is that the impact of all parameters on the overall quality of jobs in the economy ($\sigma$) can be easily determined by differentiating the function $t^{-1}(\alpha; \sigma_x, \sigma_\theta)$. Specifically, we have that
\begin{IEEEeqnarray}{rCl}
\tderiv{c_a} \sigma &=&  \frac{\partial \alpha}{\partial c_a}\deriv{\alpha}t(\alpha; \sigma_x, \sigma_{\theta})=-\frac{1}{c_l c_a^2}\frac{1}{\deriv{\sigma}t^{-1}(\sigma; \sigma_x, \sigma_{\theta})}<0 \label{eq: compstsigmaalpha} \\ %
\tderiv{\sigma_x}  \sigma & = & -\underbrace{\deriv{\sigma_x} t^{-1}(\sigma;\sigma_x, \sigma_{\theta})}_{<0}  \underbrace{\frac{1}{\deriv{\sigma}t^{-1}(\sigma; \sigma_x, \sigma_{\theta})}}_{>0}>0, \label{eq: compstsigmaA} \\
\tderiv{\sigma_{\theta}}  \sigma & =& -\underbrace{\deriv{\sigma_{\theta}} t^{-1}(\sigma;\sigma_x, \sigma_{\theta})}_{<0}  \underbrace{\frac{1}{\deriv{\sigma}t^{-1}(\sigma; \sigma_x, \sigma_\theta)}}_{>0}>0.
\end{IEEEeqnarray}
Hence, increases in the variances of skill ($\sigma_x$) and productivity ($\sigma_{\theta}$), as well as amenity provision and span-of-control costs all lead to a better overall supply of quality jobs. The intuition for these results is straightforward, although it  differs slightly across the different sources of change.

Let us start by focusing on changes in $\sigma_x, c_a$ and $c_l$, as the mechanism through which they affect the supply of jobs is similar. Consider the firms' size choices---and the resulting job distribution---in a partial equilibrium that holds constant the utility a worker of a given skill receives. Each of these three types of changes would cause all firms to hire more workers in such a partial equilibrium. However, each of these changes would also cause the larger firms to expand by more. In the case of the span-of-control and amenity provision costs, the reason is that  $c_a$ and $c_l$ govern not only the slope (the cost of expanding the workforce or amenity provision) but also the curvature of their respective cost functions; hence, relative to its initial level, a fall in $c_l$ ($c_a$) decreases the cost of workforce (amenity provision) expansion by more for larger firms. In the case of an increase in the supply of high-skilled workers, the reason is the complementarity in production between high skill workers and high productivity firms.

Of course, given that the \emph{number} of workers is fixed, it cannot be true that all firms expand in general equilibrium. Thus, the utility received by workers of all skills goes up, which increases firms' wage costs and causes a contraction of all firms. This contraction reverses the original expansion for low-productivity firms, but not for high-productivity firms. Overall, therefore, low-productivity firms contract and high-productivity firms expand; hence, the supply of quality jobs improves.

While an increase in the variance of productivity has the same final effect on the supply of quality jobs, the mechanism through which it operates is slightly different. An increase in $\sigma_\theta$ directly increases the number of quality jobs, by increasing the number of high productivity firms in existence (relative to the number of low productivity firms). Clearly, holding utility constant, this increases the demand for high skilled workers---who are mostly demanded by high-productivity firms---and decreases the demand for low-skilled workers, which translates into lower (higher) payoffs for low (high) skilled workers. As a result, in general equilibrium low productivity firms expand and high productivity firms contract. However, this second-order intensive margin effect on job quality is dominated by the first-order extensive margin increase in the number of high-productivity firms, and thus the overall supply of quality jobs improves.
\subsection{Sorting}\label{sec: sorting}
How do workers sort with firms in our model? The key to answering this question is the observation that the sorting between workers and \emph{jobs} is perfectly positive and assortative, and hence $\text{Corr}(X, H)=1$. Hence, the correlation between workers and \emph{firms} must be the same as the correlation between firms and jobs. We have already observed that the distribution of jobs within a firm is normal with mean $\theta$ and variance 1. It follows immediately that $\text{Corr}(H, \bar{\theta})=\sqrt{\frac{\text{Var}(\bar{\theta})}{\text{Var}(H)}},$
where $\bar{\theta}$ denotes the distribution of firm productivity weighted by the size of each firm.\footnote{To see why, note that a simple regression of job on firm type would result in the following relation ship between $h$ and $\theta$: $h=\theta+\epsilon$, with $\epsilon$ standard normally distributed.} It is easy to see that
$$\text{Var}(H)=\sigma^2=1+\text{Var}(\bar{\theta}),$$
so that $\text{Corr}(X, \bar{\theta})\equiv\rho=\sqrt{1-1/\sigma^2}.$
Indeed,  one can also show that
\begin{equation}
    \label{eq: corr} \text{Corr}(X^2, \bar{\theta}^2)=\rho^2=1-\frac{1}{\sigma^2}.\footnote{This can be derived using the fact that $\text{Corr}(X^2, \bar{\theta}^2)=\text{Corr}(H^2, \bar{\theta}^2)=\text{Cov}((\bar{\theta}+T)^2, \bar{\theta}^2)/\sqrt{\text{Var}(H^2) \text{Var}(\bar{\theta})},$ where $T, \bar{\theta}$ are independent and jointly normal.}
\end{equation}

Thus, the strength of the positive sorting between workers and firms depends only on the overall supply of quality jobs in the economy; the better the jobs are, the stronger the sorting.
In general, quality jobs are on offer in both high- and low-productivity firms: Even the least productive firm has some highly value-added tasks (such as management of the entire company) that need to be performed.  If high productivity firms are relatively small (or there are relatively few of them), then there are few quality jobs, but a relatively large share of those few quality jobs is supplied by low-productivity firms. If, however, highly productive firms become very large (or there are very many of them), then most jobs are of high quality, and virtually all quality jobs are supplied by high-productivity firms; however, as low-productivity firms consist primarily of low quality jobs, a sizeable proportion of bad jobs will be still supplied by low-productivity firms, even though those firms are small overall (or there is few of them). As high skill workers fill quality jobs, this implies that the higher the overall quality of jobs, the stronger the sorting between high skilled workers and high-productivity firms.

\subsection{Welfare Inequality}\label{sec: welfareinequality}
The positive relationship between average job quality and sorting is of critical importance for the relationship between overall and between-firm welfare inequality.\footnote{Of course, wage and welfare inequality coincide  in the limit when $c_a \to \infty$.} To see why, first note that the law of total variance implies that
\begin{equation} \label{eq: utilitydecomp} \text{Var}_{U}\left(\ln U\right)=\underbrace{E_{\bar{\theta}}(\text{Var}_U(\ln U|\bar{\theta}))}_{WFUI=\text{within-firm welfare inequality}} +  \underbrace{\text{Var}_{\bar{\theta}}E_U(\ln U|\bar{\theta}))}_{BFUI=\text{between-firm welfare inequality}}.\end{equation}
\cite{Bowsher2012} show that for any random variables $T$ and $Z$ we have that $\frac{\text{Var}_T(\text{E}_Z(Z| T))}{\text{Var}_Z(Z)}=\text{Corr}(Z, \text{E}_Z(Z| T))^2$. This has a simple, but profound implication for the relationship between sorting, between-firm and overall welfare inequality:
\begin{equation}
    \label{eq: bfuigen} \frac{BFUI}{\text{Var}( \ln U) }=\text{Corr}\left(\ln U, \text{E}_U(\ln U| \theta) \right)^2=\text{Corr}\left(\ln U(X), \text{E}_X(\ln U(X)| \bar{\theta}) \right)^2.
\end{equation}
Equation \eqref{eq: bfuigen} implies that the ratio of between-firm and overall welfare inequality is equal to the squared correlation between a function of skill $(\ln U(X))$ and a function of firms' type ($\text{E}_X(\ln U(X)| \theta)$). Hence, as long as these two functions are monotone, this correlation is a measure of the sorting between workers and firms. This indicates that in a wide-range of competitive models, including any generalisation of the present model that would allow for general distributions of skills, tasks or productivities, the share of overall utility inequality explained by its between-firm component will be a function of an appropriately defined measure of the strength of sorting.

In our specific case, log utility is a quadratic function of skill (by Equation \eqref{eq: utility}) and the average of squared skill is a linear function of squared productivity, with
\begin{equation}\label{eq: avgskill}
    E(X^2|\theta)=\left(\frac{\sigma_x}{\sigma}\right)^2(1+\theta^2).\footnote{As discussed above, the distribution of $H$ conditional on $\theta$ is normal with mean $\theta$ and variance 1. Thus, $H^2$ conditional on $\theta$ has a non-centralised Chi-squared distribution with 1 degree of freedom and centrality parameter $\theta^2$. Thus, $\text{E}_H(H^2 | \theta)=1+\theta^2$, and the relationship between skill and productivity follows from the assignment function $\mu(h)$.}
\end{equation}
Therefore, Equation \eqref{eq: bfuigen} simplifies to
\begin{equation}
    \label{eq: bfuispec} \frac{BFUI}{\text{Var}(\ln U) }=\text{Corr}\left(X^2, \bar{\theta}^2 \right)^2=\rho^4=\left(1- \frac{1}{\sigma^2}\right)^2.
\end{equation}
Thus, the share of overall welfare inequality explained by its between-firm component depends positively on the overall supply of quality jobs in the economy and on nothing else! Furthermore, between-firm, within-firm and overall inequality all depend just on the supply of quality jobs and the distribution of workers skills, with
\begin{IEEEeqnarray}{rCl}
   \text{Var}(\ln U)&=&0.5\sigma_x^2\sigma^2 \label{eq: varu}\\
   BFUI&=&\text{Var}(\ln U) \rho^4=0.5\sigma_x^2\sigma^2\left(1- \frac{1}{\sigma^2}\right)^2\\
    WFUI&=& \text{Var}(\ln U)(1- \rho^4)=0.5\sigma_x^2(2-\frac{1}{\sigma^2}).\footnote{$\text{Var}(\ln U)$ is computed from Equation \eqref{eq: utility}, between-firm welfare inequality follows from Equation \eqref{eq: bfuispec} and within-welfare inequality can be then computed from \eqref{eq: utilitydecomp}.}
\label{eq: WFUI}
\end{IEEEeqnarray}
Hence, overall welfare inequality and its components all increase in the supply of quality jobs; overall inequality changes by the same proportion as $\sigma^2$, between-firm inequality changes by a larger proportion than $\sigma^2$, and within-firm welfare inequality changes by a lower proportion than $\sigma^2$.

The fact that all components of welfare inequality increase in job quality in the supply of high skill workers implies, together with the discussion in Section \ref{sec: qualityjobs}, that a change in any of the four primitives increases overall welfare inequality if and only if it also increases the share of welfare inequality that is explained by between-firm welfare inequality.
This indicates that the trends in inequality observed in Norway and many other developed countries are exactly what one should expect; increases in overall welfare inequality will typically be driven by changes in between-firm rather than within-firm inequality. And while we will see that this co-movement of overall inequality and its share explained by the between-firm component is not inevitable,  once we look at wages rather than welfare or allow for changes in more than one primitive at a time, it nevertheless strikes us as remarkable that the most basic mechanism present in our economy creates such a co-movement.

The following Proposition summarises the results in Sections \ref{sec: qualityjobs} to \ref{sec: welfareinequality}.

\begin{prop}\label{prop: jobssortingwelfare}
The supply of quality jobs $\sigma^2$, sorting $\rho^2$, overall welfare inequality  $\text{Var}(\ln U)$, between-firm welfare inequality BFUI, within-firm welfare inequality WFUI and the share of overall welfare inequality explained by its between-firm component  $ \frac{BFUI}{\text{Var}(\ln U) }$ increase in $\sigma_x, \sigma_{\theta}$ and decrease in $c_a, c_l$.
\end{prop}

\subsection{Wage Inequality}\label{sec: wageinequality}
Focusing on wages makes the link between sorting, and between-firm and overall inequality less straightforward.
Let us start by decomposing overall wage inequality using the law of total variance:
\begin{equation}\label{eq: wagedecomp} \text{Var}_{W}\left(\ln W\right)=\underbrace{E_{\bar{\theta}}(\text{Var}_W(\ln W|\bar{\theta}))}_{WFWI=\text{within-firm wage inequality}} +  \underbrace{\text{Var}_{\bar{\theta}}E_W(\ln W|\bar{\theta}))}_{BFWI=\text{between-firm wage inequality}}.\end{equation}
Of course, it remains true that
\begin{equation}
    \label{eq: bfwigen} \frac{BFWI}{\text{Var}(\ln W) }=\text{Corr}\left(\ln W, \text{E}_W(\ln W| \theta) \right)^2.
\end{equation}
In contrast to utility, however, the wage a worker receives is a function of not just their own type, but also the amenities provided by the firm they work for, and the amount of effort the worker exerts. Specifically, Equations \eqref{eq: eqmwage}, \eqref{eq: FOC} and \eqref{eq: utility} imply jointly that the log wage received by a worker of skill $x$ working for firm $\theta$ is equal to:
\begin{equation}
    \label{eq: wage} \ln(w(x; a^*(\theta)))=\ln(u(x))+\ln(a^*(\theta))=\frac{\sigma}{2\sigma_x}x^2+\frac{\theta^2}{2c_a\left(\frac{\sigma}{\sigma_x}-1 \right)}+B
\end{equation}
 where $B$ is a non-stochastic term.\footnote{Specifically, $B\equiv \ln(u(0))+\ln(\frac{1}{c_a}+1)+\frac{c_l}{1+c_a}\left(\ln(\sigma_{\theta})-0.5 \ln(\sigma^2-1)+\frac{\ln(1+c_l)}{c_l}\right)$.} From this and Equation \eqref{eq: avgskill} follows immediately that the average wage paid by a firm is linear in the square of the firm's productivity  $\theta$:
 \[ \text{E}(\ln W|\theta)=0.5\frac{\sigma_x}{\sigma}\theta^2\left(1+\frac{1}{c_a \left(1-\frac{\sigma_x}{\sigma} \right)} \right)+B+0.5\frac{\sigma_x}{\sigma}.\]
Hence, while it remains true that the average wage paid by a firm is strictly increasing in the firm's type, it is not anymore true that more skilled workers \emph{always} receive a higher wage. The reason is that more productive firms pay higher wages to all of their workers. Indeed, a highly skilled worker who works for a low-productivity firm may well receive a lower wage and fewer amenities than a low skilled worker who works for a high productivity firm, because the low-productivity firm will ask its workers to exert very little effort. In terms of between-firm and overall wage inequality, this implies that the correlation between a worker's wage and the average wage paid at their firm is \emph{higher} than the correlation between their utility and the average utility of workers at their firm, with
\begin{equation}
    \label{eq: bfwispec} \frac{BFWI}{\text{Var}(\ln W) }=\frac{1}{1+\frac{\frac{1}{\rho^4}-1}{\left(1+\frac{1}{c_a \left(1-\frac{\sigma_x}{\sigma} \right) }\right)^2}}.
\end{equation}
Of course, if providing amenities is prohibitively expensive  ($c_a \to \infty$) then wage and welfare inequality coincide, and the expression above tends to $\rho^4$. If amenities are being provided, however, then the relationship between the supply of quality jobs and the share of overall wage inequality explained by its between-firm component becomes more complex.
To understand why, suppose that jobs are becoming better because more productive firms grow their workforce. Of course, the resulting increase in sorting still contributes to a more than proportional increase in between-firm wage inequality. However, there is also an additional channel present, which can dominate in some regions of the parameter space. When more productive firms expand, they must start hiring worse workers on average---both in absolute terms, and relative to the quality of workers hired by low-productivity firms. Because wages and amenities are complements in the workers' utility function, and better workers are paid higher wages (on average and within each firm), the smaller the difference in the average skill of workers hired by high- and low-productivity firms, the smaller the difference in amenities provided by those firms. However, as a fall in amenity provision causes a more than proportional decrease in the effort exertion required by the firm, smaller differences in amenities translate then into smaller compensating differentials paid by more productive firms, which decreases the correlation between individual workers' wages and the average wage paid by the firm they work for.

Overall wage inequality can be computed from Equation \eqref{eq: wage}, between-firm wage inequality can be calculated using Equation \eqref{eq: bfwispec} and within-firm inequality obtains then from \eqref{eq: wagedecomp}:
\begin{IEEEeqnarray}{rCl}
  \label{eq: varW}  \text{Var}(\ln W)&=&\text{Var}(\ln U) \left(1+\frac{\rho^4}{c_a \left(1-\frac{\sigma_x}{\sigma} \right) }\left(2+\frac{1}{c_a \left(1-\frac{\sigma_x}{\sigma} \right) } \right) \right), \\
   \label{eq: bfwi} BFWI&=&\text{Var}(\ln U)\rho^4\left( 1+\frac{1}{c_a \left(1-\frac{\sigma_x}{\sigma} \right) } \right)^2, \\
   \label{eq: wfwi} WFWI&=&WFUI.
\end{IEEEeqnarray}
Within-firm wage inequality coincides with within-firm welfare inequality because amenity provision is firm specific rather than worker-firm specific. Thus, it remains true that an increase in the supply of quality jobs increases within-firm wage inequality. However, its effect on overall inequality is ambiguous in general. %

\begin{prop}\label{prop: wageinequality}
(i) Overall wage inequality  $\text{Var}(\ln W)$, between-firm wage inequality BFWI, and the share of overall wage inequality explained by its between-firm component  $ \frac{BFWI}{\text{Var}(\ln W) }$ increase in $\sigma_x$ and decrease in $c_a$.    (ii) Further, if $\sigma_x<1$ ($\sigma_x<0.99)$ then BFWI and $\text{Var}(\ln W)$ ($\frac{BFWI}{\text{Var}(\ln W}$) increase in $\sigma_{\theta}$ and decrease in $c_l$.  If, however, $\sigma_x>1$ then BFWI, $\text{Var}(\ln W)$ and $\frac{BFWI}{\text{Var}(\ln W)}$ are all non-monotone in $\sigma_{\theta}$ and $c_l$.
\end{prop}
\deferred[proof: cl]{\subsection{Proof of Proposition \ref{prop: wageinequality}}
(i) Let us start by showing that

\begin{equation}\label{eq: two}\tderiv{c_a}c_a(1-\frac{\sigma_x}{\sigma})>0, \tderiv{\sigma_x}c_a(1-\frac{\sigma_x}{\sigma})<0.\end{equation}
The former follows as soon as we rewrite Equation \eqref{eq: inversesol} as
\begin{equation} \label{eq: carewritten}c_a(1-\frac{\sigma_x}{\sigma})=\frac{\frac{1+c_a}{c_l}\frac{\sigma_x}{\sigma}}{\frac{1}{\sigma_\theta}-\frac{1}{\sigma^2-1}}.\end{equation}
The right-hand side increases in $c_a$ both directly, and indirectly, because it is decreasing in $\sigma$ (which is, itself, decreasing in $c_a$).
To see the latter, denote $\frac{\sigma}{\sigma_x}$ by $\bar{\sigma}$, and define the function $$h(\bar{\sigma}; \sigma_x, \sigma_\theta)\equiv\left(\bar{\sigma}-1 \right)\left(\frac{1}{\sigma_\theta}-\frac{1}{\bar{\sigma}^2\sigma_x^2-1} \right).$$ As $h$ is increasing in $\bar{\sigma}$, in equilibrium it must be the case that $h^{-1}(\bar{\sigma}; A, \beta)=\bar{\sigma}$. Given this and the fact that $\deriv{\sigma_x}h>0$, it follows that $\deriv{\sigma_x} \bar{\sigma}<0$, implying that $\tderiv{\sigma_x}c_a(1-\frac{1}{\bar{\sigma}})<0$ as well.

Part (i) follows then immediately from Proposition \ref{prop: jobssortingwelfare}, and Equations \eqref{eq: varW}, \eqref{eq: bfwi}, \eqref{eq: wfwi} and \eqref{eq: two}.

(ii) %

We will start with the results regarding BFWI. Define $b \equiv \frac{\sqrt{BFUI}}{c_a(1-\frac{\sigma_x}{\sigma})}$ and note that
\[\tderiv{\sigma} b = \frac{\sqrt{2}\sigma_x\left(\sigma^2 -2\sigma_x \sigma+1 \right)}{2c_a(\sigma-\sigma_x)^2}.
\]
It is easy to see that if $\sigma_x<1$ then $\tderiv{\sigma} b>0$, and thus $\tderiv{\sigma}BFWI>0$. Hence, by Proposition \ref{prop: jobssortingwelfare}, if $\sigma_x<1$ then BFWI increases in $\sigma_\theta$ and decreases in $c_a$.

Suppose, instead, that $\sigma_x>1$. In that case,  $\tderiv{\sigma} b>0$ if and only if $\sigma> \sigma_x+\sqrt{\sigma_x^2-1}$, which implies that
$\tderiv{\sigma}BFWI>0$ if $\alpha>\sqrt{1-\frac{1}{\sigma_x^2}}\left(\sigma_\theta^{-1}-1/\left(2\left(\sigma_x^2-\sigma_x\sqrt{\sigma_x^2-1}-1 \right)\right)\right).$ In other words, BFWI decreases in $c_l$ for $c_l$ smaller than $(1+\frac{1}{c_a})\left(\sqrt{1-\frac{1}{\sigma_x^2}}\left(\sigma_\theta^{-1}-1/\left(2\left(\sigma_x^2-\sigma_x\sqrt{\sigma_x^2-1}-1 \right)\right)\right)\right)^{-1}$ and increases in $\sigma_\theta$ for $\sigma_\theta> \left(\alpha/\sqrt{1-\frac{1}{\sigma_x^2}}+
1/\left(2\left(\sigma_x^2-\sigma_x\sqrt{\sigma_x^2-1}-1 \right)\right) \right)^{-1}$. The fact that  $\tderiv{\sigma} b>0$ if and only if $\sigma> \sigma_x+\sqrt{\sigma_x^2-1}$ implies also that $\lim_{\sigma \to \sigma_x}\tderiv{\sigma} b=- \infty$, as the denominator of $\tderiv{\sigma} b$ goes to zero as $\sigma \to \sigma_x$. One can easily infer from Equation \eqref{eq: inversesol} that $\sigma \to \sigma_x$ as $c_l \to \infty$ or $\sigma_\theta \to 0$. As
\begin{IEEEeqnarray*}{rCl}
    \tderiv{\sigma} BFWI&=&2(\tderiv{\sigma} \sqrt{BFUI}+\tderiv{\sigma} b )\sqrt{BFUI} \\
    \tderiv{\sigma} \text{Var}(\ln W)&=&\tderiv{\sigma} WFWI +\tderiv{\sigma}BFWI,
\end{IEEEeqnarray*}
and both $\tderiv{\sigma} \sqrt{BFUI}$ and $\tderiv{\sigma} WFWI$ have finite limits as $\sigma \to \sigma_x$, it follows that BFWI and $\tderiv{\sigma} \text{Var}(\ln W)$ decrease in $\sigma_\theta$ for very small $\sigma_\theta$, and increase in $c_l$ for very large values of $c_l$.\footnote{A careful reader may be surprised at this result. It is quite clear that for $\sigma_\theta=0$ the model reduces to a \cite{Costrell2004} model, in which between-firm wage inequality is equal 0. How can it thus be that BFWI decreases for low $\sigma_\theta$? The reason is that if $\sigma_x>1$ then the equilibrium $\sigma$ is discontinuous at $\sigma_\theta=0$. For $\sigma_\theta=0$, we have that $\sigma=1$; however, for any $\sigma_\theta>0$, $\sigma$ is strictly greater than $\sigma_x$. We find this discontinuity fascinating, but not of first-order importance, as $\sigma_x>1$ would imply a much higher level of within-firm wage inequality than those observed in the data.}

To show the results regarding  $\frac{BFWI}{\text{Var}W}$ let us define $e\equiv\frac{\rho^2}{\sqrt{1-\rho^4}}\frac{1}{1-\sqrt{1-\rho^2}\sigma_x}$, and note that
\[\tderiv{\rho^2} e=\frac{\left(\rho^6+\rho^2-2\right) \sigma_x+2 \sqrt{1-\rho^2}}{2 (\rho^2-1)^2 (\rho^2+1)^{3/2} \left(\sqrt{1-\rho^2} \sigma_x-1\right)^2}
.\]
One can show that for $\sigma_x<2/(2.24(2\sqrt{5}/5))<1$ the numerator of this expression is always positive. Thus, if $\sigma_x<0.99$ then $\tderiv{\rho^2} e>0$. As we can write
$$\frac{BFWI}{\text{Var}W}=\frac{1}{1+\frac{1}{\left(\frac{1}{\sqrt{\frac{1}{\rho^4}-1}}+e\right)^2}}=\frac{1}{1+\frac{WFWI}{BFWI}}$$
it follows by Proposition \ref{prop: jobssortingwelfare} that if $\sigma_x<0.99$, then $\frac{BFWI}{\text{Var}W}$ increases in $\sigma_\theta$ and decreases in $c_l$. Next, note that because $\tderiv{\sigma} WFWI >0$, it follows that if $\tderiv{\sigma} BFWI <0$ then $\tderiv{\sigma} \frac{BFWI}{\text{Var}W} <0$. Thus, it follows from the discussion above that if $\sigma_x>1$, then $\frac{BFWI}{\text{Var}W}$ decreases in $\sigma_\theta$ for very small $\sigma_\theta$, and increases in $c_l$ for very large values of $c_l$. Finally, it is easy to see that $\tderiv{\rho^2} e>0$ if an only if $\sigma_x<\frac{1}{\sqrt{1-\rho^2}(\rho^4+\rho^2+2)}$; for any value of $\sigma_x$, there must exist some values of $\rho$ that are close enough to $1$ such that this condition is satisfied. It is easy to see by inspection of Equations \eqref{eq: inversesol} and \eqref{eq: corr} that if $\sigma_\theta \to \infty$ or $c_l \to 0$, then $\rho \to 1.$  Overall, therefore, if $\sigma_x>1$, then $\frac{BFWI}{\text{Var}W}$ is non-monotone in both $c_l$ and $\sigma_\theta$, as required.
}
Hence, changes in the distribution of workers' skill and the cost of amenity provision affect between-firm, within-firm and overall welfare and wage inequality in the same direction.  Changes in firms' type distribution and span-of-control cost, can---in theory---affect wage and welfare inequality differently in certain regions of the parameter space. Similarly, in theory it is possible that, say, a decrease in the span-of-control cost may increase wage inequality but decrease the share of wage inequality that is explained by its between-firm component.

Critically, however, the condition $\sigma_x<0.99$ is extremely likely to be satisfied in practice: Virtually all calibrations of the model to real-world data would estimate $\sigma_x$ to be lower than $0.99$. To see why, recall that $\sigma>1$ and note that Equations \eqref{eq: WFUI} and \eqref{eq: wfwi} imply that
\[\sigma_x<\frac{\sqrt{2}\sqrt{WFWI}}{\sqrt{2-\frac{1}{\sigma^2}}}<\sqrt{2}\sqrt{WFWI}.\]
For developed countries, the empirical estimates of within-firm wage inequality are well below $0.7$, implying that $\sigma_x$ would indeed take values lower than $0.99$ if calibrated to match this moment.\footnote{\cite{Tomaskovic-Devey2020-dp} report overall and between-firm wage inequality for 14 developed countries (Table S3.3), with the largest value of overall wage variance being 0.585 (Canada in 2007). In our data, within-firm wage inequality in Norway peaks in 2008 at 0.144.}
 This indicates that, in practice, any change in a single primitive of our model changes overall wage inequality and its share explained by the between-wage component in the same direction.

\subsection{Differential Impact of Changes in Primitives} \label{sec: diffimpact}
The preceding discussion implies that, in practice, overall wage inequality and its share explained by the between-firm component may \emph{not} co-move only if two or more changes in primitives are counteracting each other.

As an example, suppose that the variance of skill $\sigma_x$ and the cost of amenity provision have both decreased at the same time, so that these two changes affect all outcomes in opposite directions. Suppose further, that their impacts on overall welfare inequality exactly offset each other. In that case, the ratio of welfare inequality explained by its between-firm component would increase, because for a decrease in $\sigma_x$ and $c_a$ to have an effect on welfare inequality of the same magnitude, the fall in $c_a$ must have had a stronger effect on the equilibrium supply of quality jobs and thus also on sorting. One can similarly, if much more laboriously, show that overall wage inequality would also increase in such a scenario, as would its share explained by the between-firm component.

\begin{prop}\label{prop: relativeimpact}
For a given set of parameters $\sigma_x, \sigma_{\theta}, c_a, c_l$ define two scalars $y_{\sigma_x}, y_{c_l}$ such that $$\tderiv{\sigma_x} \text{Var}(\ln U)=y_{c_a}\tderiv{c_a} \text{Var}(\ln U)=y_{c_l}\tderiv{c_l} (\ln U).\footnote{We omit $\sigma_{\theta}$ as its impact on the outcomes considered in this Proposition must trivially be the same as the impact of $c_l$.}$$

Then the relative (local) impact of changes in $\sigma_x, c_a, c_l$ on $\frac{BFUI}{\text{Var}(\ln U)}$, $ \frac{BFUI}{\text{Var}(\ln U)}$ and  $\text{Var}(\ln W)$ is as follows:
\begin{IEEEeqnarray*}{rrCl}
  \text{(i)} \qquad &  y_{c_a}\tderiv{c_a} \frac{BFUI}{\text{Var}(\ln U)}&=&y_{c_l}\tderiv{c_l} \frac{BFUI}{\text{Var}(\ln U)}>\tderiv{\sigma_x} \frac{BFUI}{\text{Var}(\ln U)}\\
  \text{(ii)}\qquad &   y_{c_a}\tderiv{c_a} \frac{BFWI}{\text{Var}(\ln W)}&>&\max\{\tderiv{\sigma_x} \frac{BFWI}{\text{Var}(\ln W)},y_{c_l}\tderiv{c_l} \frac{BFWI}{\text{Var}(\ln W)}\}\\
 \text{(iii)} \qquad &     y_{c_a}\tderiv{c_a} \text{Var}(\ln W)&>&\max\{\tderiv{\sigma_x}\text{Var}(\ln W),y_{c_l}\tderiv{c_l}\text{Var}(\ln W)\}.
\end{IEEEeqnarray*}
Furthermore, if $\text{Var}(\ln U)<0.5$ then
\begin{IEEEeqnarray*}{rrCl}
   \text{(iv)} \qquad & y_{c_l}\tderiv{c_l} \frac{BFWI}{\text{Var}(\ln W)}&>&\tderiv{\sigma_{x}} \frac{BFWI}{\text{Var}(\ln W)},\\
  \text{(v)} \qquad &  y_{c_l}\tderiv{c_l}\text{Var}(\ln W)&>&\tderiv{\sigma_x}\text{Var}(\ln W).
\end{IEEEeqnarray*}
\end{prop}

\deferred[proof: relativeimpact]{\subsection{Proof of Proposition \ref{prop: relativeimpact}}
Throughout this proof, we will denote $\text{Var}(\ln U)$ by $V$.

(i) Note that sorting can be rewritten as
\begin{equation}\label{eq: rhorewritten}
    \rho^2=1-\frac{\sigma_x^2}{2V},
\end{equation}
which immediately yields that
\[y_{c_a}\tderiv{c_a} \rho^4=y_{c_l}\tderiv{c_l} \rho^4=4y_{c_l}\tderiv{c_l}V\frac{\sigma_x^2}{2V^2}\rho^2>4(\tderiv{\sigma_x}V\frac{\sigma_x^2}{2V^2}-\frac{\sigma_x}{2V})\rho^2=\tderiv{\sigma_x} \rho^4,\]
and part (i) follows from Equation \eqref{eq: bfuispec}.

(ii) Define $\delta \equiv \frac{1}{c_a(1+\sigma_x/\sigma)}$. Using Equations \eqref{eq: carewritten} and \eqref{eq: rhorewritten} we can rewrite it as
\[\delta=\left(\frac{\frac{1+c_a}{c_l}\frac{\frac{\sqrt{2}}{2}\sigma_x^2}{\sqrt{V}}}{\frac{1}{\sigma_\theta}-\frac{1}{\frac{2V}{\sigma_x^2}-1}}\right)^{-1}.\]
Hence, conditioning on $V$, $\delta$  decreases in $c_a$ and $\sigma_x$, and increases in $c_l$.
The ratio of between-firm to overall wage inequality can be readily rewritten as
\[
    \frac{BFWI}{\text{Var}(\ln W)}=\frac{1}{1+(\frac{1}{(1-\frac{\sigma_x^2}{2V})^2}-1)\left(1+\delta \right)^{-2}}.
\]
Again, conditioning on $V$, this expression decreases in $c_a$ and $\sigma_x$, and increases in $c_l$. Thus, it follows that a fall in $c_a$ that has the same impact on $V$ as an increase  in $\sigma_x$ (decrease in $c_l$) will increase $\frac{BFWI}{\text{Var}(\ln W)}$ by more.

(iii) The expression for $\text{Var}(\ln W)$ can be rewritten as
\[\text{Var}(\ln W)=V\left (1+\delta(1-\frac{\sigma_x^2}{2V})^2(2+\delta)\right).\]
Clearly, this expression also decreases in $c_a$ and $\sigma_x$, and increases in $c_l$, and thus part (iii) follows by the same reasoning as above.

(iv) Let us now define $\bar{\delta}\equiv \frac{1-\frac{1}{\sigma^2}}{c_a(1-\sigma_x/\sigma)},$
and note that
\[\bar{\delta}=\frac{1-\frac{1}{\sigma^2}}{c_a(1-\frac{\sqrt{2V}}{\sigma^2})}=c_a^{-1}\left(1+\frac{\sqrt{2V}-1}{\sigma^2-\sqrt{2V}}\right)=c_a^{-1}\left(1+\frac{\sqrt{2V}-1}{\frac{2V}{\sigma_x^2}-\sqrt{2V}}\right).\] \
Hence, if $V<0.5$  and conditioning on $V$, then $\bar{\delta}$ decreases in $\sigma_x$ and does not depend on $c_l$. We can then clearly write
\[\frac{BFWI}{\text{Var}(\ln W)}=\frac{1}{1+(1-(1-\frac{\sigma_x^2}{2V})^2)\left(1-\frac{\sigma_x^2}{2V}+\bar{\delta} \right)^{-2}},\]
so that if $V<0.5$  (and conditioning on $V$) then $\frac{BFWI}{\text{Var}(\ln W)}$  decreases in $\sigma_x$ and does not depend on $c_l$, and part (iv) follows.

(v) Finally, we can rewrite the expression for $\text{Var}(\ln W)$ as
\[\text{Var}(\ln W)=V(1+\bar{\delta}(2(1-\frac{\sigma_x^2}{2V})+\bar{\delta}))\]
from which it follows immediately that (conditioning on V), if $V<0.5$ then $\text{Var}(\ln W)$ decreases in $\sigma_x$ but does not depend on $c_l$; hence, part (v) follows as well.
}
 Proposition \ref{prop: relativeimpact} normalises the size of changes in the primitives by requiring that each affects overall welfare inequality equally, and then compares their impact on selected other outcomes. The impact on the ratio of between-firm to overall welfare inequality is the most straightforward one. Because $c_l$ and $c_a$ affect overall welfare inequality only through job quality/sorting, and $\sigma_x$ affects it directly as well, if $c_l/c_a$ have the same impact on overall welfare inequality as $\sigma_x$ then their impact on sorting must be greater than that of $\sigma_x$.

 The relative impact that different changes in primitives have on overall wage inequality and on its share that is explained by the between-firm component is less obvious. It should be fairly clear why changes in the cost of amenity provision have the strongest impact, as $c_a$ has both a strong effect on sorting and impacts $\text{Var}(\ln W)$ ($\frac{BFWI}{\text{Var}(\ln W)}$) directly. In general, it is ambiguous whether changes in the skill distribution or changes in the span-of-control cost have greater impact on overall wage inequality and on its share that is explained by the between-firm component; on the one hand, the increase in $\sigma_x$ has a direct positive effect on $\text{Var}(\ln W)$ ($\frac{BFWI}{\text{Var}(\ln W)}$) through $1/(1-\sigma_x/\sigma)$, on the other hand, the increase in $\sigma_x$ has a much weaker impact on sorting/supply of quality jobs (compared to a fall in $c_l$ that has equal impact on overall welfare inequality).
 However, one can show quite easily that if $\text{Var}(\ln U)<0.5$ then the impact of a fall in the span-of-control cost must be stronger; to see why, let us rewrite Equation \eqref{eq: varW}:
 \[\text{Var}(\ln W)=\text{Var}(\ln U)\left(1+\frac{1-\frac{1-\sqrt{2\text{Var}(\ln U)}}{\sigma^2-\sqrt{2\text{Var}(\ln U)}}}{c_a} \right)\left(2(1-\frac{1}{\sigma^2})+\frac{1-\frac{1-\sqrt{2\text{Var}(\ln U)}}{\sigma^2-\sqrt{2\text{Var}(\ln U)}}}{c_a} \right).\]
Partialling out the impact of changes in  $\text{Var}(\ln U)$, $\text{Var}(\ln W)$ increases in  $\sigma$  if  $\text{Var}(\ln U)<0.5$; and because a fall in $c_l$ increases $\sigma$ by more, its impact on overall wage variance must be greater as well. The ratio of between-firm to overall wage inequality can be rewritten analogously.

Critically, again, overall wage inequality is smaller than $0.5$ in the vast majority of the 14 developed countries studied in \cite{Tomaskovic-Devey2020-dp}, with the only exceptions being the US, Israel and Canada. Most importantly, our estimate of overall wage inequality in Norway peaks at 0.217 in 2007. Therefore, we expect that in our calibration changes in the span-of control cost will have a stronger impact on overall
wage inequality (relative to their impact on overall welfare inequality) than changes in the variance of skill.

\section{Calibration Exercise}\label{sec: calibration}
We first describe the data used in the calibration exercise (Section \ref{ssec: data}), then identify the model (Section \ref{sec: id}), calibrate it (Section \ref{sec: cal}), and finally simulate counterfactual scenarios with an eye on establishing the main drivers of the observed changes the composition of inequality (Section \ref{sec: counter}).

\subsection{Data}\label{ssec: data}
The main data used for calibration are matched employer-employee data in Norway, drawn from administrative registers maintained by Statistics Norway \citep{ssb2020-ameld,ssb2020-tab,ssb2020-atmlto}.
The data contain the universe of workers and firms recorded in the registers. %
Each worker is assigned a unique person identifier, which allows us to track workers over time.
Similarly, each firm is identified by a unique firm identifier that does not change over time.
When a worker matches to a firm in a given year, an entry for the worker-firm pair is created for that year. %
For each worker-firm-year combination, we have details regarding the worker's date of birth, the firm's industry sector, and crucially, the worker's wage at the firm, along with other attributes. %
Since we observe all workers within a firm, as well as all firms in the economy, we can reliably compute firm sizes and determine wage distributions for each year.

Earnings records for workers pertain to remuneration, which include
fixed salary, holiday pay, sickness and maternity benefits, and other cash benefits paid to the worker deemed as remuneration by the Norwegian Tax Administration \citep{skd2018-kode}. %
All earnings are adjusted to 2013 values using the Consumer Price Index published by Statistics Norway \citep{ssb2023-cpi}.\footnote{We chose 2013 as the reference year following \citet{Song2018}.}

Employment and wage data prior to 2015 are available on an annual basis.
Statistics Norway links the employee register with individual tax files to provide wages and labour information. %
Since 2015, employment statistics are reported on a monthly basis. Employers submit salary and employment information directly to the employee register, the tax office, and Statistics Norway \citep{ssb2020-regi}.
To ensure wages are measured at the same frequency, we aggregate monthly earnings within each year, starting from 2015, to generate annual earnings.
Due to this change in the reporting scheme in 2015 \citep{ssb2014-change}, we divide our sample into two parts, 1995 to 2014, and 2015 to 2019.\footnote{In our baseline sample, we do not observe structural breaks in key statistics such as the average number of workers, number of firms, and average wage, from 2014 to 2015.  Nonetheless, as a precautionary measure, we divide the sample period into pre-2014 and post-2015 segments.}
What constitutes a firm in our sample? We consider businesses listed in the Central Register of Establishments and Enterprises. %
A business is an entity that primarily operates within a specific industrial classification \citep{ssb2023-virk}. %
It is the level at which employers submit work and pay information for tax purposes (see, for instance, a walk-through at \citet{skd2023-ameld}). %
A business is a subdivision of the legal unit, an enterprise.
As businesses and enterprises can be consistently identified from 1995 in our data \citep[e.g.,][]{ssb2020-arb}, we start our sample period from 1995. %
We conclude our sample period in 2019 since it is the last year unaffected by the COVID-19 pandemic.

Our analytic sample consists of individuals aged between 20 and 60, following \citet{Song2018}.
To identify which workers are weakly attached to the labour force or to a certain job, we introduce the concept of Basic Amount (``Grunnbeløp'') \citep[e.g.,][]{skd2023-gpj}. %
The Basic Amount is updated each year by the Parliament.
It is the legal amount at which the national insurance scheme becomes applicable.\footnote{The Basic Amount is similar to CPI in concept. It is is updated according to wage growth, while CPI is calculated on the price of consumer goods and services.
The two do not always align. From 1995 to 2019, the Basic Amount has grown by 155\%, whereas CPI has only grown by 65\%.}  %
We remove individuals who earn less than two times the Basic Amount in that year, following common practice \citep[e.g.,][]{Markussen2019-gc,Hoen2022-sf}.\footnote{On average, Norwegians earn approximately six times the Basic Amount.} %
Regarding firms, we remove firms with less than 5 employees because it is not meaningful to calculate within-firm variance for extremely small firms.
We exclude firms and workers in the following sectors: public administration, defence, and social security schemes subject to public administration; education; paid work in private households; and international organisations and bodies. Wages in these sectors are often flat according to preset wage schedules.
These restrictions result in a sample of 1.3 million workers and 54,000 firms per year on average.
Throughout the sample period, the average annual growth rate in the number of workers is 1.5\%, and that for the number of firms is 2.5\%.

\subsection{Identification}\label{sec: id}
Let us denote the vector of exogenous parameters by $\Sigma\equiv(\sigma_x, \sigma_{\theta}, c_a, c_l, A)$. We will identify $\Sigma$ using five moments of the wage distribution:

\begin{enumerate}
    \item within-firm wage inequality ($\text{E}_{\bar{\theta}}\left(\text{Var}{W}(\ln W | \bar{\theta}) \right)$),
    \item the variance of the within-firm wage variances of wages ($\text{Var}_{\bar{\theta}}\left(\text{Var}{W}(\ln W | \bar{\theta}) \right)$),
    \item the unweighted within-firm wage inequality ($\text{E}_{\theta}\left(\text{Var}{W}(\ln W | \theta) \right)$),
    \item variance of log wages ($\text{Var}_{W} (\ln(W))$),
    \item and the expected (log) wage in the economy ($\text{E}_{W}\left (\ln (W)\right)$).
\end{enumerate}

Of course, with just 5 parameters the model is very much over-identified. We chose these 5 particular moments because amenity and effort exertion (both of which are not fully observed in the data) do not affect the variance of wages within any firm; and thus, the first three moments depend on the cost of amenity provision parameter $c_a$ only through its impact on $\alpha \equiv (1+1/c_a)/c_l$ which itself affects only the equilibrium supply of quality jobs $\sigma^2$.\footnote{A sensible alternative to using one of these 3 moments would be to estimate an AKM-like fixed effects regression, and identify $\sigma$ from the correlation between the worker and firm fixed effects. We have decided against it, as there are many modelling choices that go into an AKM regression and such moment would be significantly more `processed'. However, in Section \ref{sec: cal} we will compare the AKM covariance estimates found in the literature to those resulting from our calibration as an non-targeted moment, to validate the quality of calibration. }
The first three moments can be thus used to identify $\sigma_x, \sigma_{\theta}$ and $\alpha$; the cost of amenity provision is then identified from the overall wage variance, which it affects directly---together with $\alpha$, this identifies $c_l$; finally, the TFP parameter $A$ is identified from the average (log) wage in the economy.

The variance of the within-firm wage variances of wages and the unweighted within-firm wage inequality are equal to
\begin{IEEEeqnarray}{rCl}
   \label{eq: varvar} \text{Var}_{\bar{\theta}}\left(\text{Var}{W}(\ln W | \bar{\theta}) \right)&=&\left(\sqrt{2}\sigma_x^2\left(1-\frac{1}{\sigma^2} \right)\right)^2,\\
   \label{eq: unWFWI} \text{E}_{\theta}\left(\text{Var}{W}(\ln W | \theta) \right)&=&0.5\left(\frac{\sigma_x}{\sigma}\right)^2(1+2\sigma_{\theta}^2).\footnote{To see how these expressions are derived, recall that the jobs offered by firm $\theta$ have a normal distribution with mean $\theta$ and variance 1 (see footnote following Equation \eqref{eq: avgskill}). It follows from the properties of non-centralised Chi-squared distribution that the variance of $H^2$ offered by firm $\theta$ is equal to $\text{Var}_H(H^2 | \theta)=2+4\theta^2$. Hence, from the facts that $X^2= \left(\frac{\sigma_x}{\sigma}H \right)^2$ and $\ln W=\frac{\sigma}{2\sigma_x}X^2+\frac{\theta^2}{2c_a\left(\frac{\sigma}{\sigma_x}-1 \right)}+B$ in equilibrium it follows that $\text{Var}_W(\ln(W) | \theta)=0.5(\frac{\sigma_x}{\sigma})^2(1+2\theta^2)$. Finally, $\text{E}_{\theta}(\theta^2)=\sigma_\theta^2$ and $\text{Var}_{\bar{\theta}}(\theta^2)=2(\sigma^2-1)^2$ from which Equations \eqref{eq: varvar} and \eqref{eq: unWFWI} follow.}
\end{IEEEeqnarray}
Equations \eqref{eq: varvar} and \eqref{eq: wfwi} jointly identify the equilibrium supply of quality jobs, with
\begin{equation}\label{eq: sigmaident}
  \sigma^2=\frac{\sqrt{2}WFWI-0.5\sqrt{\text{Var}_{\bar{\theta}}\left( \text{Var}_X (\ln (W) | \bar{\theta}) \right)}}{\sqrt{2}WFWI-\sqrt{\text{Var}_{\bar{\theta}}\left( \text{Var}_X (\ln (W) | \bar{\theta}) \right)}}.
\end{equation}
Observe that the right-hand side of this equation increases in the variance of within firm wage dispersion; hence, to rationalise a large variance of within-firm wage dispersion in our economy, the equilibrium supply of quality jobs must be large as well. %
This is because $\sigma$ is small only if most jobs are offered by firms with low $\theta^2$, which are all similar to each other. Therefore,  if $\sigma$ is small and we weight the population by firm size, then the variance of within firm wage variance does not differ much across firms.

Once we know $\sigma^2$, the variance of skill (high productivity firms) can be identified from Equations \eqref{eq: wfwi} (and \eqref{eq: unWFWI}):
\begin{IEEEeqnarray}{rCl}
   \label{eq: sigmaxident} \sigma_x^2&=&\frac{2WFWI}{2-\frac{1}{\sigma^2}},\\
   \label{eq: sigmathetaident} \sigma_{\theta}^2&=&0.5\left(\frac{ \text{E}_{\theta}\left(\text{Var}{W}(\ln W | \theta) \right)}{WFWI}(2\sigma^2-1)-1\right).
\end{IEEEeqnarray}
Within-firm inequality depends on $\sigma$ and $\sigma_x$ only, and thus, for a given $\sigma$, the variance of skill must be high for the model to rationalise a larger degree of within-firm inequality. The ratio of within firm wage inequality \emph{unweighted} by firm size to within firm wage inequality \emph{weighted} by firm size depends only on $\sigma$ and $\sigma_{\theta}$. Weighted within-firm inequality can be large compared to unweighted inequality, only if there are few high productivity firms, but those firms are large.

With $\sigma, \sigma_x$ and $\sigma_{\theta}$ identified, $\alpha$ follows immediately from the equilibrium condition (Equation \eqref{eq: eqmsigma}). The cost of amenity provision $c_a$ can then be easily identified from overall wage inequality, which it affects directly, with
\begin{equation}
  \label{eq: caident} c_a=\frac{\left(\frac{\sqrt{2 \text{Var}_{W} (\ln(W))-WFWI}}{\sigma_x \sigma(1-\frac{1}{\sigma^2})}-1\right)^{-1}}{1-\frac{\sigma_x}{\sigma}}.
\end{equation}
Essentially, the value of $c_a$ in our calibration is determined by the gap between the actual value of overall wage inequality and the value that the model produces without the presence of amenities. The larger this gap is, the lower the cost of providing amenities must be.

The span-of-control cost is then given by $\alpha$ and $c_a$:
\begin{equation}
  \label{eq: clident} c_l=\frac{1+\frac{1}{c_a}}{\left(\frac{\sigma}{\sigma_x}-1\right) (\frac{1}{\sigma_{\theta}^2}-\frac{1}{\sigma^2-1})}.
\end{equation}
The span of control cost is, effectively, calibrated as a remainder: It is the value of $c_l$ that ensures that the calibrated $\sigma$ equalibrates the model given the estimates of all the other parmaters.

Finally, note that---by Equation \eqref{eq: wage}---the average (log) wage in the economy equals
\[E_W(\ln(W))=0.5\sigma_x\left(\sigma \left(1+\frac{1}{c_a(1-\frac{\sigma_x}{\sigma})} \right) -\frac{1}{c_a(\sigma-\sigma_x)} \right)+B,\]
which can be used to identify the TFP parameter $A$ using Equation \eqref{eq: u0} and the definition of $B$.

\subsection{Calibration and Simulation}\label{sec: cal}

The targeted moments are calculated using the Norwegian administrative data for the years 1995 to 2019, as discussed in Section \ref{ssec: data}. Instead of using the raw moments, we produce 1500 samples for each year using a bootstrap-like resampling procedure, and then calculate the moments and calibrate the model separately for each of these samples; the reported central estimates of our moments and parameters simply represent the average value over the 1500 samples, and the confidence intervals correspond to the 2.5th and 97.5th percentiles of the 1500 estimates.
Furthermore, in the case of the  variance of within firm variances we de-noise the raw estimate. The raw estimate overestimates the \emph{true} variation in the variances of within-firm wage distributions because it is also affected by the variance of the within-firm variance estimator; as some firms in our sample are small, this bias can be substantial.\footnote{To understand the source of this bias better, suppose that all firms were of the same type, and thus with an infinite population would hire the same distribution of workers. As our sample is finite, the firms would nevertheless differ in the type of workers they end up hiring in the data, and thus the observed variance of wages would differ across firms, even though the `true' variance of variances is 0.} We estimate the size of this bias and subtract this estimate from the raw data moment of the variance of within-firm wage variances. The details of both our resampling procedure and the adjustment applied to the variance of within-firm variances are described in Online Appendix \ref{app: bootstrap}.
\begin{figure}[!tb]
\begin{center}
\subfloat{\resizebox{\textwidth}{!}{\input{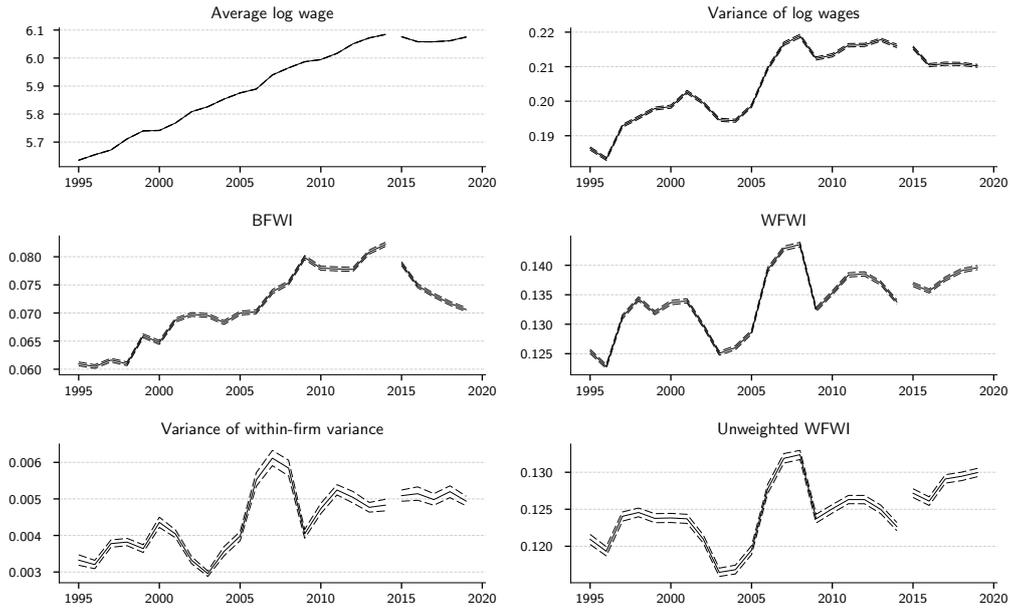}}}
\caption{Moments used in calibration}
\label{fig: moments}
\end{center}

\footnotesize{\label{page: figcont} Notes: Plot of the data moments and confidence used for the calibration. Top panel: (1) Mean log wage; (2) Variance of log wages. Middle panel: (1) The across firms, firm-size weighted variance of (within firm) mean log wages; (2) The across firms, firm-size weighted mean  of the variance of log wages. Bottom panel: (1) The across firm, firm-size weighted variance of the (within-firm) variance of log wages; (2) The across firms, size-unweighted mean  of the variance of log wages. Both the moments and the confidence intervals were obtained through the resampling procedure discussed in detail in Online Appendix \ref{app: bootstrap}.}

\end{figure}

The targeted moments are reported in Figure \ref{fig: moments}. Three observations are in order. First, from 1995 to 2014 overall wage inequality and both of its components have increased, but the increases in BFWI were larger in both absolute and relative terms. Second, since 2015 both overall wage variance and between-firm wage inequality have been decreasing, the latter quite sharply; however, within-firm wage inequality has continued to rise. Third, changes in the variance of within-firm wage variance have mimicked closely the changes in WFWI between 1995 and 2014; since 2014, however, the variance of within-firm wage variance has remained constant.

\begin{figure}[!tb]
\begin{center}
\subfloat{\resizebox{\textwidth}{!}{\input{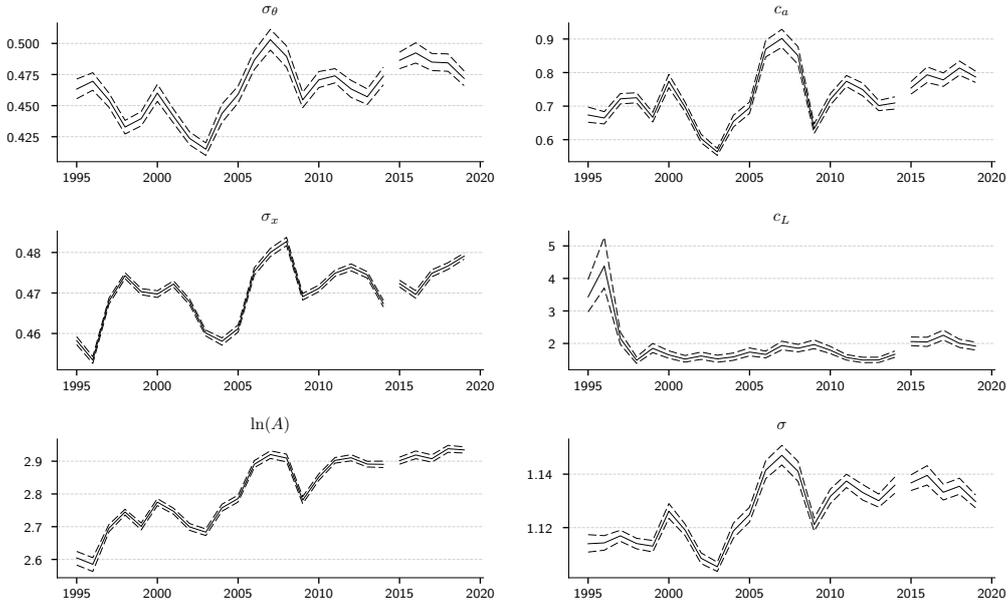}}}
\caption{Calibrated parameters}
\label{fig: paramstime}
\end{center}
\footnotesize{Notes: The calibrated parameters from 1995 to 2019. Top panel: (1) the variance of productivity $\sigma_\theta$, (2) cost of amenity provision $c_a$. Middle panel: (1) variance of skill $\sigma_x$, (2) span-of-control cost $c_l$. Bottom panel: (1) log of total factor productivity $\ln A$, (2) equilibrium supply of quality jobs $\sigma$. The dashed lines depict the 95\% confidence intervals.}

\end{figure}

The calibrated parameters are graphed in Figure \ref{fig: paramstime}.\footnote{The change in the estimates from year 1995 (for estimates from 1996 to 2014) and 2015 (for estimates from 2016 to 2019), together with the corresponding confidence intervals, is depicted in Figure \ref{fig: diffstime} in the Online Appendix.} All targeted moments are matched exactly within each of the 1500 year-specific samples, and thus also on average. The calibration reveals that the variance of productivity and the cost of providing amenities remained relatively stable between 1995 and 2014; the change from 1995 to 2014 is insignificant in the case of $\sigma_{\theta}$ and only marginally significant in the case of $c_a$. After 2015, the variance of productivity seems to have fallen slightly, and the cost of amenity provision has slightly increased.

The changes in the remaining parameters are large and statistically significant. The variance of skill has reached its peak in 2007, but has remained significantly higher than its 1995 level in 2014; this increase in $\sigma_x$ has continued after 2015. The span-of-control cost plummeted between 1996 and 1998, but has remained relatively stable since. Finally, the TFP of the Norwegian economy has been steadily climbing between 1995 and 2014, but has levelled off since 2015.

The changes in $\sigma_\theta, \sigma_x, c_a$ and $c_l$ have caused an increase in the equilibrium supply of quality jobs $\sigma$ between 1995 and 2014 (with a peak in 2007).\footnote{The fact that a number of our calibrated parameters peaks in 2007 is an artefact of the fact that our calibration interprets any observed changes in inequality as caused by a change in the primitives; thus, the spike in within-firm variance just before the financial crisis, which was likely caused by rents in reality, our model interprets as being caused by changes in $\sigma_\theta, \sigma_x$ and $c_a$. This is partly why, when discussing the results, we focus on the 1995-2014 comparison, rather than the evolution of the parameters throughout the entire period.} Since 2015, however, the supply of quality jobs has actually decreased in the Norwegian economy.

\begin{figure}[!tb]
\begin{center}
\subfloat{\resizebox{\textwidth}{!}{\input{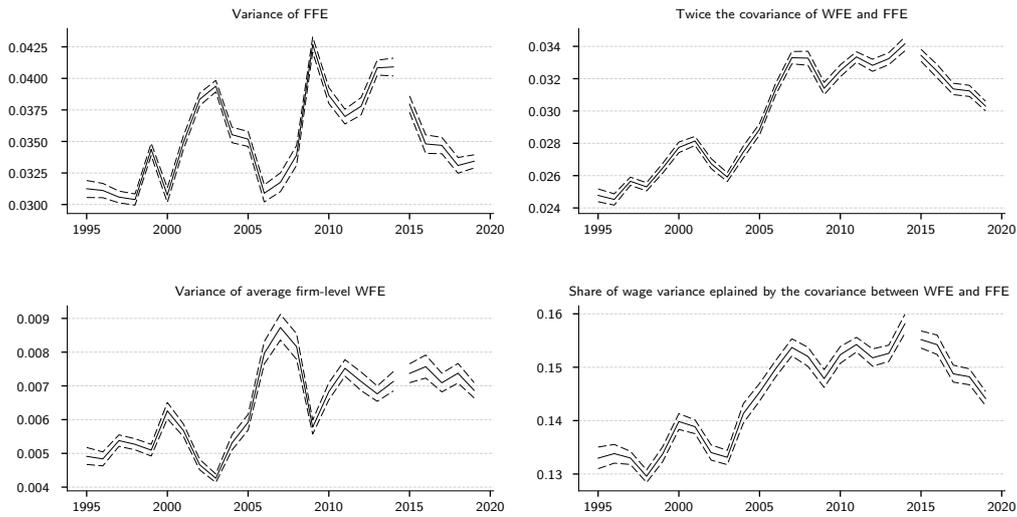}}}
\caption{Decomposition of between-firm wage inequality}
\label{fig: FEdecomp}
\end{center}

\footnotesize{\label{page: figcont} Notes:  Components of between-firm wage inequality from 1995 to 2019. Top panel: (1) variance of firm fixed effects, (2) twice the covariance of worker fixed effects and firm fixed effects. Bottom panel: (1) variance of average firm-level worker fixed effects, (2) the ratio of twice the covariance between worker fixed effects and firm fixed effects to overall wage variance. The dashed lines depict the 95\% confidence intervals. }

\end{figure}

We use our calibrated parameter values to simulate a number of moments that are not directly observed in the data. Most importantly, we use the calibration to decompose between-firm wage inequality into the variance of the compensating differential for effort, the variance of average within-firm worker utility, and the covariance between compensating differentials and utility (Figure \ref{fig: FEdecomp}). This exercise is analogous to decomposing BFWI into the variance of firm fixed effects (FFE), firm-level average worker fixed effects (WFE), and twice the covariance between these two \citep[see Equation (5) and Table IV in][]{Song2018}. The reason is that, as a direct consequence of Equation \eqref{eq: wage}, in our economy a worker's utility is exactly equal to that worker's fixed effect in an AKM regression; and the compensating differential paid by a firm is exactly equal to both the amount of amenities provided by the firm and the firm's fixed effect in an AKM regression. See Online Appendix \ref{sec: WFEFFE} for a discussion on how to perform the AKM variance decomposition in our model.

 Our simulation indicates that between 1995 and 2014 all three components of between-firm wage inequality have increased. In relative terms, the changes in all components are very similar in magnitude. In absolute terms, the increases in the variance of firm fixed effects and twice the covariance between worker and firm fixed effects are of similar magnitude, and dwarf the changes in the variance of average firm-level worker fixed effects. The (non-targeted) estimates of the variance of firm fixed effects and the covariance between worker and firm fixed effects can be used to validate our calibration, by comparing them to empirical estimates from the literature. Specifically, \cite{Bonhomme2023} report the results of an AKM fixed effects regression in Norway for the years 2009--2014 (Figure F.10). After correcting for the limited mobility bias, their estimates (the FE-HE, FE-HO and CRE estimators) of the ratio of twice the covariance between firm- and worker-fixed-effects and overall annual wage variance ranges between 13\%--17\%. In our calibration, the central simulation of the same moment for the years 2009--2014 ranges between 14.8\% and 15.8\%. Since this moment was not targeted by us, we find it reassuring that it is in a similar range to that reported in the literature. More worryingly, however, \cite{Bonhomme2023} report that the variance of  firm fixed effects explained between 9\% and 10\% of overall annual wage variance in Norway between 2009 and 2014, whereas in our calibrated model it explains between 17\% and 20\%, which is closer to the uncorrected estimates from \cite{Bonhomme2023}.\footnote{Admittedly, \cite{Bonhomme2023} use a different sample of firms and individuals. For example, they use full-time equivalent wages to measure labour market attachment.}   %

\begin{figure}[!tb]
\begin{center}
\subfloat{\resizebox{\textwidth}{!}{\input{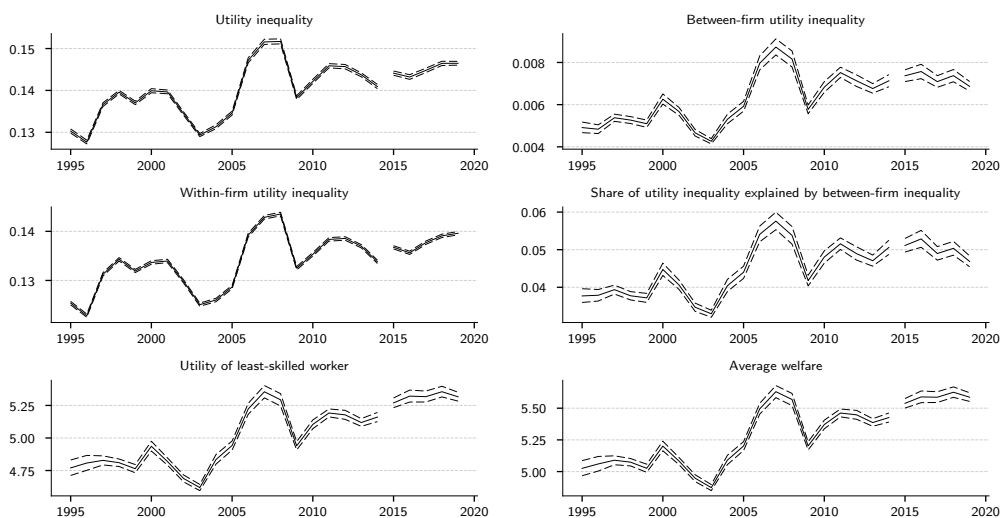}}}
\caption{Decomposition of utility inequality}
\label{fig: utdecomp}
\end{center}

\footnotesize{\label{page: figcont} Notes:  Components of utility inequality from 1995 to 2019. Top panel: (1) utility inequality, (2) between-firm utility inequality. Middle panel: (1) within-firm utility inequality, (2) share of utility inequality explained by between-firm inequality. Bottom panel: (1) utility of least-skilled worker, (2) average welfare. The dashed lines depict the 95\% confidence intervals.}

\end{figure}

 Of the moments depicted in Figure \ref{fig: FEdecomp}, the between-firm differences in utility/WFE are particularly interesting, as---within our model---they correspond exactly to between-firm utility inequality. This allows us then to compute overall utility inequality---by adding between-firm utility inequality to within-firm wage inequality---and decompose it into its within- and between-firm components (Figure \ref{fig: utdecomp}). Interestingly, between-firm utility inequality was responsible for less than 9\% of overall utility inequality throughout our sample.\footnote{This is, perhaps, not so surprising, given our assumption that all differences in wages paid to workers of identical skill are compensating differentials. However, given the findings in \cite{Sorkin2018}, this assumption may not be drastically off.} The changes in utility inequality from 1994 to 2014 were broadly in the same direction as the changes in wage inequality, with between-firm welfare inequality responsible for 65\% of the overall increase in welfare inequality  over this time-frame. The magnitudes, however, are quite different. As we report in Table \ref{tab: counter} below, overall wage inequality increased by 0.03, whereas overall utility inequality increased by only 0.011; thus, only a little more than a third of the overall increase in wage inequality reflects an actual increase in welfare inequality. As we have learnt from Figure \ref{fig: FEdecomp} and the corresponding discussion, the remaining two thirds of the increase in wage inequality reflect (in equal measure) an increase in the variance  in the compensating differential for effort exertion and twice the covariance between workers' utility and the compensating differentials.

Strikingly, between 2015 and 2019 the trends in overall wage and welfare inequality have diverged: The decrease in overall wage inequality masks a continued increase in overall welfare inequality. In the broadest of sense,  the reason for this divergence of trends is the fact that between-firm welfare inequality is much smaller than between-firm wage inequality in our calibrated model, whereas within-firm wage and welfare inequalities are always the same. Thus, the continued increase in within-firm welfare inequality dominates the fall in between-firm welfare inequality, but the fall in between-firm wage inequality dominates the increase in within-firm wage inequality. Of course, this raises a deeper question: Why are within- and between-firm inequality changing in opposite direction since 2015; we will attempt to answer this question in Section \ref{sec: counter} below.

 The last two  panels Figure \ref{fig: utdecomp} depict the changes in (a) the utility of the worst-off worker and (b) the average welfare in the economy, both of which were much higher in 2014 than in 1995, and have remained stable after 2015. Thus, the technological changes responsible for the observed increases in inequality have, in fact, made everyone better off.

\subsection{Counterfactual Analysis}\label{sec: counter}
In this section, we will decompose the changes in wage and welfare inequality into the share explained by the change in each of the four parameters. However, because the impact that a change in any parameter has on the outcomes depends on the other parameters, the order of change matters for the result: The impact of a change in, say, $\sigma_x$ may be drastically different if all the other parameters are at their 1995 levels than if the other parameters are at their 2014 levels. We deal with this issue by simulating the effect of changes in parameters for each of the 24 permutations of the order in which the vector  $(\sigma_\theta, \sigma_x, c_a, c_l)$ may have changed between any two points in time, and then averaging over these permutations.

\begin{table}[!htbp]
\centering
\begin{scriptsize}
\resizebox{\textwidth}{!}{\begin{tabular}{C{1cm}C{1.9cm}*{5}{C{1.6cm}}}
     \toprule
&& $ \sigma_{\theta} $ & $ \sigma_x $ & $ c_a $ & $ c_l $ &Overall \\
   \midrule
\multirow{8}{=}{VarW}& \multirow{4}{=}{1995-2014\\ \tiny{conf. interval}  \\ share explained \\ \tiny{conf. interval} } & \multirow{4}{=}{\centering 0.007 \\ \tiny{[-0.001,0.015]} \\ 24.1 \\ \tiny{[-1.7,49.0]}} & \multirow{4}{=}{\centering 0.010 \\ \tiny{[0.009,0.012]} \\ 34.6 \\ \tiny{[29.9,39.4]}} & \multirow{4}{=}{\centering -0.006 \\ \tiny{[-0.011,-0.001]} \\ -20.8 \\ \tiny{[-37.3,-3.7]}} & \multirow{4}{=}{\centering 0.018 \\ \tiny{[0.015,0.022]} \\ 62.2 \\ \tiny{[50.9,72.9]}} & \multirow{4}{=}{\centering 0.030 \\ \tiny{[0.029,0.030]} \\ 100.0 \\ \tiny{[100.0,100.0]}} \\
 &&&&& \\
 &&&&& \\
 &&&&& \\
 &\multirow{4}{=}{2015-2019 \\ \tiny{conf. interval}  \\ share explained \\ \tiny{conf. interval} } & \multirow{4}{=}{\centering -0.010 \\ \tiny{[-0.017,-0.004]} \\ 194.3 \\ \tiny{[73.9,312.3]}} & \multirow{4}{=}{\centering 0.008 \\ \tiny{[0.007,0.009]} \\ -150.1 \\ \tiny{[-185.4,-117.8]}} & \multirow{4}{=}{\centering -0.005 \\ \tiny{[-0.009,-0.001]} \\ 99.5 \\ \tiny{[23.3,181.4]}} & \multirow{4}{=}{\centering 0.002 \\ \tiny{[-0.001,0.005]} \\ -43.7 \\ \tiny{[-99.8,12.3]}} & \multirow{4}{=}{\centering -0.005 \\ \tiny{[-0.006,-0.005]} \\ 100.0 \\ \tiny{[100.0,100.0]}} \\
 &&&&& \\
 &&&&& \\
 &&&&& \\
 \hline\multirow{8}{=}{WFWI}& \multirow{4}{=}{1995-2014 \\ \tiny{conf. interval}  \\ share explained \\ \tiny{conf. interval} } & \multirow{4}{=}{\centering 0.001 \\ \tiny{[-0.000,0.002]} \\ 10.9 \\ \tiny{[-0.8,22.1]}} & \multirow{4}{=}{\centering 0.005 \\ \tiny{[0.005,0.006]} \\ 62.2 \\ \tiny{[54.0,70.4]}} & \multirow{4}{=}{\centering -0.000 \\ \tiny{[-0.000,-0.000]} \\ -1.3 \\ \tiny{[-2.3,-0.2]}} & \multirow{4}{=}{\centering 0.002 \\ \tiny{[0.002,0.003]} \\ 28.3 \\ \tiny{[22.8,33.7]}} & \multirow{4}{=}{\centering 0.008 \\ \tiny{[0.008,0.009]} \\ 100.0 \\ \tiny{[100.0,100.0]}} \\
 &&&&& \\
 &&&&& \\
 &&&&& \\
 &\multirow{4}{=}{2015-2019 \\ \tiny{conf. interval}  \\ share explained \\ \tiny{conf. interval} } & \multirow{4}{=}{\centering -0.001 \\ \tiny{[-0.002,-0.001]} \\ -49.8 \\ \tiny{[-85.4,-17.0]}} & \multirow{4}{=}{\centering 0.004 \\ \tiny{[0.003,0.005]} \\ 142.4 \\ \tiny{[120.1,167.7]}} & \multirow{4}{=}{\centering -0.000 \\ \tiny{[-0.000,-0.000]} \\ -3.8 \\ \tiny{[-6.5,-1.0]}} & \multirow{4}{=}{\centering 0.000 \\ \tiny{[-0.000,0.001]} \\ 11.1 \\ \tiny{[-3.1,25.7]}} & \multirow{4}{=}{\centering 0.003 \\ \tiny{[0.002,0.003]} \\ 100.0 \\ \tiny{[100.0,100.0]}} \\
 &&&&& \\
 &&&&& \\
 &&&&& \\
 \hline\multirow{8}{=}{BFWI}&\multirow{4}{=}{1995-2014 \\ \tiny{conf. interval}  \\ share explained \\ \tiny{conf. interval} } & \multirow{4}{=}{\centering 0.006 \\ \tiny{[-0.000,0.013]} \\ 29.3 \\ \tiny{[-2.1,59.8]}} & \multirow{4}{=}{\centering 0.005 \\ \tiny{[0.004,0.006]} \\ 23.6 \\ \tiny{[20.3,27.2]}} & \multirow{4}{=}{\centering -0.006 \\ \tiny{[-0.011,-0.001]} \\ -28.5 \\ \tiny{[-51.8,-5.0]}} & \multirow{4}{=}{\centering 0.016 \\ \tiny{[0.013,0.019]} \\ 75.6 \\ \tiny{[61.6,89.2]}} & \multirow{4}{=}{\centering 0.021 \\ \tiny{[0.021,0.022]} \\ 100.0 \\ \tiny{[100.0,100.0]}} \\
 &&&&& \\
 &&&&& \\
 &&&&& \\
 &\multirow{4}{=}{2015-2019 \\ \tiny{conf. interval}  \\ share explained \\ \tiny{conf. interval} } & \multirow{4}{=}{\centering -0.009 \\ \tiny{[-0.014,-0.003]} \\ 109.5 \\ \tiny{[40.4,177.5]}} & \multirow{4}{=}{\centering 0.004 \\ \tiny{[0.003,0.005]} \\ -47.7 \\ \tiny{[-55.6,-39.7]}} & \multirow{4}{=}{\centering -0.005 \\ \tiny{[-0.009,-0.001]} \\ 62.8 \\ \tiny{[15.5,110.1]}} & \multirow{4}{=}{\centering 0.002 \\ \tiny{[-0.001,0.005]} \\ -24.6 \\ \tiny{[-56.0,6.8]}} & \multirow{4}{=}{\centering -0.008 \\ \tiny{[-0.009,-0.008]} \\ 100.0 \\ \tiny{[100.0,100.0]}} \\
 &&&&& \\
 &&&&& \\
 &&&&& \\
 \hline\multirow{8}{=}{$ \frac{\text{BFWI}}{\text{Var}W} $}&\multirow{4}{=}{1995-2014 \\ \tiny{conf. interval}  \\ share explained \\ \tiny{conf. interval} } & \multirow{4}{=}{\centering 0.018 \\ \tiny{[-0.001,0.037]} \\ 34.2 \\ \tiny{[-2.4,69.7]}} & \multirow{4}{=}{\centering 0.007 \\ \tiny{[0.006,0.008]} \\ 12.7 \\ \tiny{[10.7,14.8]}} & \multirow{4}{=}{\centering -0.019 \\ \tiny{[-0.034,-0.003]} \\ -35.5 \\ \tiny{[-64.8,-6.3]}} & \multirow{4}{=}{\centering 0.047 \\ \tiny{[0.039,0.055]} \\ 88.6 \\ \tiny{[72.0,104.4]}} & \multirow{4}{=}{\centering 0.054 \\ \tiny{[0.052,0.055]} \\ 100.0 \\ \tiny{[100.0,100.0]}} \\
 &&&&& \\
 &&&&& \\
 &&&&& \\
 &\multirow{4}{=}{2015-2019 \\ \tiny{conf. interval}  \\ share explained \\ \tiny{conf. interval} } & \multirow{4}{=}{\centering -0.025 \\ \tiny{[-0.040,-0.009]} \\ 84.3 \\ \tiny{[30.7,137.6]}} & \multirow{4}{=}{\centering 0.005 \\ \tiny{[0.004,0.006]} \\ -17.5 \\ \tiny{[-20.2,-14.8]}} & \multirow{4}{=}{\centering -0.015 \\ \tiny{[-0.027,-0.004]} \\ 52.0 \\ \tiny{[13.0,91.2]}} & \multirow{4}{=}{\centering 0.006 \\ \tiny{[-0.002,0.013]} \\ -18.8 \\ \tiny{[-43.1,5.3]}} & \multirow{4}{=}{\centering -0.030 \\ \tiny{[-0.031,-0.028]} \\ 100.0 \\ \tiny{[100.0,100.0]}} \\
 &&&&& \\
 &&&&& \\
 &&&&& \\
 \hline\multirow{8}{=}{VarU}&\multirow{4}{=}{1995-2014 \\ \tiny{conf. interval}  \\ share explained \\ \tiny{conf. interval}} & \multirow{4}{=}{\centering 0.001 \\ \tiny{[-0.000,0.003]} \\ 13.7 \\ \tiny{[-1.1,27.0]}} & \multirow{4}{=}{\centering 0.006 \\ \tiny{[0.005,0.006]} \\ 52.2 \\ \tiny{[43.7,61.1]}} & \multirow{4}{=}{\centering -0.000 \\ \tiny{[-0.000,-0.000]} \\ -1.7 \\ \tiny{[-2.9,-0.3]}} & \multirow{4}{=}{\centering 0.004 \\ \tiny{[0.003,0.004]} \\ 35.8 \\ \tiny{[28.6,43.0]}} & \multirow{4}{=}{\centering 0.011 \\ \tiny{[0.010,0.011]} \\ 100.0 \\ \tiny{[100.0,100.0]}} \\
 &&&&& \\
 &&&&& \\
 &&&&& \\
 &\multirow{4}{=}{2015-2019 \\ \tiny{conf. interval}  \\ share explained \\ \tiny{conf. interval}} & \multirow{4}{=}{\centering -0.002 \\ \tiny{[-0.004,-0.001]} \\ -103.9 \\ \tiny{[-204.3,-29.9]}} & \multirow{4}{=}{\centering 0.004 \\ \tiny{[0.004,0.005]} \\ 188.4 \\ \tiny{[134.8,266.9]}} & \multirow{4}{=}{\centering -0.000 \\ \tiny{[-0.000,-0.000]} \\ -7.4 \\ \tiny{[-11.5,-2.4]}} & \multirow{4}{=}{\centering 0.001 \\ \tiny{[-0.000,0.001]} \\ 22.9 \\ \tiny{[-6.0,57.2]}} & \multirow{4}{=}{\centering 0.002 \\ \tiny{[0.002,0.003]} \\ 100.0 \\ \tiny{[100.0,100.0]}} \\
 &&&&& \\
 &&&&& \\
 &&&&& \\
 \hline\multirow{8}{=}{BFUI}&\multirow{4}{=}{1995-2014 \\ \tiny{conf. interval}  \\ share explained \\ \tiny{conf. interval} } & \multirow{4}{=}{\centering 0.001 \\ \tiny{[-0.000,0.001]} \\ 24.0 \\ \tiny{[-2.2,44.2]}} & \multirow{4}{=}{\centering 0.000 \\ \tiny{[0.000,0.000]} \\ 14.1 \\ \tiny{[10.4,19.0]}} & \multirow{4}{=}{\centering -0.000 \\ \tiny{[-0.000,-0.000]} \\ -3.0 \\ \tiny{[-4.8,-0.7]}} & \multirow{4}{=}{\centering 0.001 \\ \tiny{[0.001,0.002]} \\ 64.9 \\ \tiny{[48.6,84.7]}} & \multirow{4}{=}{\centering 0.002 \\ \tiny{[0.002,0.003]} \\ 100.0 \\ \tiny{[100.0,100.0]}} \\
 &&&&& \\
 &&&&& \\
 &&&&& \\
 &\multirow{4}{=}{2015-2019 \\ \tiny{conf. interval}  \\ share explained \\ \tiny{conf. interval} } & \multirow{4}{=}{\centering -0.001 \\ \tiny{[-0.001,-0.000]} \\ 188.8 \\ \tiny{[127.7,304.1]}} & \multirow{4}{=}{\centering 0.000 \\ \tiny{[0.000,0.000]} \\ -65.0 \\ \tiny{[-164.1,-34.9]}} & \multirow{4}{=}{\centering -0.000 \\ \tiny{[-0.000,-0.000]} \\ 20.5 \\ \tiny{[1.8,81.5]}} & \multirow{4}{=}{\centering 0.000 \\ \tiny{[-0.000,0.000]} \\ -44.3 \\ \tiny{[-137.6,15.9]}} & \multirow{4}{=}{\centering -0.001 \\ \tiny{[-0.001,-0.000]} \\ 100.0 \\ \tiny{[100.0,100.0]}} \\
 &&&&& \\
 &&&&& \\
 &&&&& \\
 \hline\multirow{8}{=}{$ \frac{\text{BFUI}}{\text{Var}U} $}&\multirow{4}{=}{1995-2013 \\ \tiny{conf. interval}  \\ share explained \\ \tiny{conf. interval}} & \multirow{4}{=}{\centering 0.004 \\ \tiny{[-0.000,0.007]} \\ 26.6 \\ \tiny{[-2.6,48.3]}} & \multirow{4}{=}{\centering 0.000 \\ \tiny{[0.000,0.001]} \\ 3.8 \\ \tiny{[2.6,5.3]}} & \multirow{4}{=}{\centering -0.000 \\ \tiny{[-0.001,-0.000]} \\ -3.3 \\ \tiny{[-5.2,-0.8]}} & \multirow{4}{=}{\centering 0.009 \\ \tiny{[0.008,0.011]} \\ 72.9 \\ \tiny{[53.5,98.4]}} & \multirow{4}{=}{\centering 0.013 \\ \tiny{[0.010,0.016]} \\ 100.0 \\ \tiny{[100.0,100.0]}} \\
 &&&&& \\
 &&&&& \\
 &&&&& \\
 &\multirow{4}{=}{2015-2019 \\ \tiny{conf. interval}  \\ share explained \\ \tiny{conf. interval} } & \multirow{4}{=}{\centering -0.005 \\ \tiny{[-0.009,-0.002]} \\ 128.7 \\ \tiny{[84.8,170.2]}} & \multirow{4}{=}{\centering 0.000 \\ \tiny{[0.000,0.001]} \\ -10.9 \\ \tiny{[-19.4,-7.2]}} & \multirow{4}{=}{\centering -0.000 \\ \tiny{[-0.001,-0.000]} \\ 12.4 \\ \tiny{[1.6,38.1]}} & \multirow{4}{=}{\centering 0.001 \\ \tiny{[-0.000,0.003]} \\ -30.2 \\ \tiny{[-77.5,9.8]}} & \multirow{4}{=}{\centering -0.004 \\ \tiny{[-0.007,-0.002]} \\ 100.0 \\ \tiny{[100.0,100.0]}} \\
 &&&&& \\
 &&&&& \\
 &&&&& \\
 \bottomrule \end{tabular}}
\end{scriptsize}
\caption{The average impact of changes in $\sigma_{\theta}, \sigma_x, c_a$ and $c_l$ on 7 outcomes of choice. \\ \begin{scriptsize}Notes: All effects are calculated by calibrating the model to each draw of our resampling procedure; the headline effects are then just the average across the 1500 draws for each year, whereas the confidence intervals are calculated by ranking the outcomes from lowest to highest and reporting the 0.025 and 0.975 percentiles.\end{scriptsize}\label{tab: counter}}
\end{table}

Table \ref{tab: counter} reports how the changes from 1995 to 2014 and from 2015 to 2019 in each of the four main primitives of our models affected 7 outcomes of interest: overall wage inequality, within-firm wage inequality, between-firm wage inequality, the ratio of between-firm to overall wage inequality, overall welfare inequality, between-firm welfare inequality, and the ratio of between-firm to overall welfare inequality. Starting with the 1995 to 2014 adjustments, the bulk of the overall increase in overall wage and welfare inequality can be attributed to changes in $\sigma_x$ and the span-of-control cost $c_l$; this is unsurprising given that the changes in the other two parameters were, at best, marginally significant. Having said that, note that changes in the distribution of high productivity firms ($\sigma_{\theta}$) nevertheless explain a non-trivial share of the change in between-firm wage and welfare inequality (29\% in the case of BFWI and 24\% in the case of BFUI). Changes to the cost of amenity provision had a significantly \emph{negative} effect on between-firm wage inequality, which almost exactly offset the impact of changes in $\sigma_{\theta}$; however, changes in $c_a$ have little impact on any outcomes that do not depend on $c_a$ directly, that is on within-firm inequality and any type of welfare inequality.

The fall in the span-of-control cost $c_l$ increased welfare inequality by 0.004, whereas the increase in the variance of skill/skill-biased technological change parameter $\sigma_x$ raised welfare inequality by 0.06; overall, the change in $\sigma_x$ explained the majority of the change in welfare inequality. However, as we have learnt in Section \ref{sec: diffimpact}, conditioning on their impact on welfare inequality, changes in the span-of-control cost have a greater impact on wage inequality than changes in $\sigma_x$, provided that overall welfare inequality is as low as in Norway. Thus, it is the change in the span-of-control cost that has caused the bulk of the increase in overall wage inequality (0.018 or 62\%), with the increase in $\sigma_x$ contributing just 35\% (0.01 in absolute terms).  In other words, over three quarters of the increase in wage inequality caused by the fall in the span-of-control cost reflected changes in the compensating differential for effort; in the case of the increase in the supply of high-skilled workers, the majority (60\%) of the increase in wage inequality reflects actually changes in welfare inequality, with only 40\% being caused by changes to compensating differentials.

The final aspect of the adjustments between 1995 and 2014 that is worth noting, is the fact that the change in $\sigma_x$ is responsible for just 3.8\% of the overall increase in the share of overall inequality that is explained by its between-firm component. Recall from Section \ref{sec: welfareinequality} that this ratio is equal to $\rho^4$, which itself is a function of only the equilibrium supply of quality jobs; hence, the change in $\sigma_x$ has a minimal impact on $\sigma$, and thus the vast majority of its large impact on other outcomes is direct.

From 2015 to 2019, we can observe large impact of changes in all primitives except perhaps for the span-of-control cost $c_l$. As discussed in Section \ref{sec: calibration}, overall wage inequality has decreased after 2015, but overall welfare inequality continued its climb. Interestingly, the impact of changes in $\sigma_x$ especially (but also in $c_l$ to a smaller degree) remained broadly the same as in the pre-2014 period. What has changed, is that post-2015 the inequality increases caused by $\sigma_x$ and $c_l$ have been counteracted by a decrease in all inequality components spurned by an increase in $c_a$ and a fall in $\sigma_\theta$. However, as we have discussed at length in Section \ref{sec: diffimpact}, relative to their impact on welfare inequality changes in $\sigma_x$ have a weak effect on wage inequality and on sorting. Thus, while the continued increase in the variance of skill was enough to ensure that welfare inequality continued to climb, the fall in the variance of productivity and the increase in the cost of amenity provision have caused a fall in wage inequality and sorting.

\section{Concluding Remarks}\label{sec: conrem}
This paper develops a novel, extremely tractable model of workers' sorting within and across firms, and uses the model to shed some light on the drivers of the observed changes in overall, within- and between-firm wage inequality in Norway from 1995 to 2019.

The tractability of the model allows us to derive a comprehensive set of comparative statics results with respect to changes in four primitives: the variance of skill, the variance of productivity, the cost of amenity provision and the span-of-control cost. The main theoretical insight is the existence of a firm link between changes in overall wage inequality, and its share explained by the between-firm component. Specifically, we find that---for realistic parameter values---if a change in one of our four primitives increases overall  wage inequality, it also must increase the ratio of between-firm to overall wage inequality. While it is possible for overall and between-firm inequality to not co-move, if changes in two or more primitives counteract each other, it appears that their co-movement should be treated as the default behaviour, rather than a surprise.

We then use our model to analyse the causes of changes in wage inequality between 1995 and 2019 in Norway. We document that overall and between-firm inequality have indeed co-moved in Norway throughout this period; prior to 2014 both were increasing, and after 2015 both were decreasing. After calibrating the model to the Norwegian data, we find that welfare inequality has been evolving qualitatively similarly to wage inequality before 2014, but post 2015 welfare inequality continued to increase. Given that changes in any of our four primitives can match the observed changes in wage inequality qualitatively, we perform a counterfactual exercise to find which of these four potential changes was quantitatively responsible for the observed changes. For 1995 to 2014, we find that a decrease in the span-of-control cost drove most of the change in observed wage inequality, while the increase in the variance of skill was responsible for the majority of the increase in welfare inequality.

\appendix
\section{Omitted Proofs and Derivations}\label{sec: omm}
\subsection{Derivation of Equation \ref{eq: density}}\label{app: mixture}
First, define $p=1+\frac{1}{\sigma_{\theta}}-\frac{\alpha}{\frac{\sigma}{\sigma_x}-1}$. Using this and Equation \eqref{eq: size}, Equation \eqref{eq: densityorig} can be rewritten as
\begin{IEEEeqnarray*}{rCl}
f(h)&=&\frac{L^*(0)}{2 \pi \sigma_{\theta} } \cdot \int_{-\infty}^{\infty} \exp \left(\frac{1+\frac{1}{\sigma_{\theta}}-p}{2} \theta^{2}-\frac{1}{2}\left(h- \theta\right)^{2}-\frac{1}{2}\left (\frac{\theta}{\sigma_{\theta}}\right )^{2}\right) \, d {\theta} \\
&=&\frac{L^*(0)}{2 \pi \sigma_{\theta} } \cdot \int_{-\infty}^{\infty} \exp \left(-\frac{1}{2}\left[\theta^{2}p-2 h \theta+h^{2}\right]\right) \, d {\theta} \\
&=&\frac{L^*(0)}{2 \pi \sigma_{\theta} } \cdot \int_{-\infty}^{\infty} \exp \left(-\frac{1}{2 }\left(\theta \sqrt{p} -\sqrt{\frac{1}{p}} h\right)^2-\frac{1}{2 } h^{2}\frac{p-1}{p}\right) \, d {\theta} \\
&=&\frac{L^*(0)}{\sigma_{\theta} \sqrt{p}} \cdot  \phi  \left (h \sqrt{1-\frac{1}{p}} \right )  \underbrace{\int_{-\infty}^{\infty} \sqrt{p} \phi \left(- \sqrt{p}\left(\theta -\frac{1}{p} h\right)\right) \, d {\theta}}_{=1},
\end{IEEEeqnarray*}
which is trivially equal to Equation \ref{eq: density}.
\shownow{proof: Aca}
\shownow{proof: cl}
\shownow{proof: beta}
\shownow{proof: relativeimpact}

\let\oldthebibliography\thebibliography
\renewcommand{\thebibliography}[1]{%
  \oldthebibliography{#1}%
  \setlength{\itemsep}{2pt}%
  \setstretch{1}%
}

  \bibliographystyle{chicago}
  \bibliography{unified}

\clearpage  %

\Large ONLINE APPENDIX

 \renewcommand{\thesection}{\Alph{section}}
 \renewcommand{\thecor}{\arabic{cor}}
  \renewcommand{\theprop}{\arabic{prop}}
  \setcounter{theo}{0}
   \renewcommand{\thetheo}{OA.\arabic{theo}}
\setcounter{section}{1}
\renewcommand{\theequation}{OA.\arabic{equation}}
\setcounter{equation}{0}
\renewcommand{\thetable}{OA.\arabic{table}}
\setcounter{table}{0}
\renewcommand{\thefigure}{OA.\arabic{figure}}
\setcounter{figure}{0}

\normalsize
\section{Firm and Worker Fixed Effects}\label{sec: WFEFFE}
Since \cite{Abowd1999} a large empirical literature has focused on estimating worker fixed effects (WFE) and firm fixed effects (FFE) using matched employer-employee data sets, and then using the estimated effects to perform various variance decomposition exercises.  As a direct consequence of Equation \eqref{eq: wage}, in our economy the worker fixed effect estimated through any AKM-like regressions would be exactly equal to the worker's utility, and the firm fixed effects would be exactly equal to the firm's amenity provision. The standard decomposition of overall wage variance into the variance of firm fixed effects, worker fixed effects and twice the covariance between these two types of fixed effects takes the following form in our model
\begin{equation}
    \label{eq: variancedecomp1} \text{Var}(\ln W)= \underbrace{0.5(\sigma_x \sigma)^2}_{\text{Variance of WFE}}+\underbrace{0.5\left(\frac{\sigma_x \sigma \rho^2}{c_a(1-\frac{\sigma_x}{\sigma})}\right)^2}_{\text{Variance of FFE}}+\underbrace{\frac{\left(\sigma_x \sigma \rho^2\right)^2}{c_a(1-\frac{\sigma_x}{\sigma})}.}_{2*\text{covariance between WFE and FFE}}
\end{equation}

The proportion of overall wage variance explained by (twice) the covariance of worker and firm fixed effects ($2*\text{Cov}(WFE, FFE)/\text{Var}(\ln W)$) is of particular interest in this literature, and is used to measure the contribution of the sorting between high wage workers and high wage firms to overall wage inequality. Indeed, this ratio of such importance, that it is often referred to as simply `sorting' and changes in this ratio are referred to as `changes in sorting' \citep[see, for example, ][]{Song2018, Bonhomme2023}. An advantage of our simple model is that it clarifies that conflating changes in  $2*\text{Cov}(WFE, FFE)/\text{Var}(\ln W)$ with changes in sorting is not always warranted.

To see why, note that the correlation between $X^2$ and $\bar{\theta}^2$ ($\rho^2$) is clearly a correct measure of the strength of sorting in our model, because it captures only the actual interdependence between true worker and firm types.\footnote{Of course, other measures of interdependence between $X^2$ and $\bar{\theta}^2$, such as Kendell's tau or Spearman's rho, could be also used to correctly measure the strength of sorting.} One can see immediately from Equation \eqref{eq: variancedecomp1} that $2*\text{Cov}(WFE, FFE)/\text{Var}(\ln W)$ will not, in general, be equal to $\rho^2$. However, the problem runs deeper, because \emph{the share of variance explained by the covariance between worker and firm fixed effects can change in the opposite direction than the correlation between worker and firm types does.} To keep the resulting expressions simple, let us write out  the inverse of $2*\text{Cov}(WFE, FFE)/\text{Var}(\ln W)$:
\begin{IEEEeqnarray*}{rCl}
    \frac{\text{Var}(\ln W)}{2*\text{Cov}(WFE, FFE)}=1+\frac{0.5}{\rho^2}\left(\sqrt{\frac{\text{Var}(WFE)}{\text{Var}(FFE)}}+\sqrt{\frac{\text{Var}(FFE)}{\text{Var}(WFE)}} \right),
\end{IEEEeqnarray*}
which, in our model, reduces to
\[\frac{\text{Var}(\ln W)}{2*\text{Cov}(WFE, FFE)}=1+0.5\frac{c_a(1-\sqrt{1-\rho^2}\sigma_x)}{\rho^4}+0.5\frac{1}{c_a(1-\sqrt{1-\rho^2}\sigma_x)}.\]
The third term decreases in $\rho^2$, and hence contributes to $2*\text{Cov}(WFE, FFE)/\text{Var}(\ln W)$ changing in the same direction as $\rho^2$. The second term, however, can increase in $\rho^2$ for $\rho^2$ sufficiently close to 1. This indicates that the \emph{empirical measure} of sorting decreases in the \emph{theoretically correct measure} of sorting if $c_a$ is large but finite, and the true degree of sorting in the economy is large. Note that in these circumstances both the variance of firm fixed effects and the ratio  $2*\text{Cov}(WFE, FFE)/\text{Var}(\ln W)$ would be low, which is consistent with many empirical estimates.
Fortunately, this issue can be easily remedied by using the \emph{correlation} of firm and worker fixed effects to measure the strength of sorting. As worker fixed effects are linear in $x^2$ and firm fixed effects are linear in $\theta^2$ in our economy, this correlation is equal to $\rho^2$. Of course, any value of this correlation that lies between 0 and 1 can be explained by our model.

\cite{Song2018} provide a further decomposition of between-firm wage inequality into  the variance of the firm fixed effect, twice the covariance between the worker and firm fixed effects and the variance of the average worker fixed-effect within each firm. In our model, this last component is clearly equal to between-firm utility inequality, as workers within the same firm receive the same amenities. Hence, in our economy the \cite{Song2018} decomposition reduces to
\begin{IEEEeqnarray*}{rCl}
    BFWI&=&\underbrace{0.5\left(\frac{\sigma_x \sigma \rho^2}{c_a(1-\frac{\sigma_x}{\sigma})}\right)^2}_{\text{Variance of FFE}}+\underbrace{\frac{\left(\sigma_x \sigma \rho^2\right)^2}{c_a(1-\frac{\sigma_x}{\sigma})}}_{2*\text{covariance between WFE and FFE}}+ \underbrace{0.5\rho^4\left(\sigma_x \sigma \rho^2\right)^2}_{\text{Variance of average WFE=BFUI.}}
\end{IEEEeqnarray*}

\section{Random sampling}\label{app: bootstrap}

The confidence intervals reported in all tables and figures from Sections \ref{sec: cal} and \ref{sec: counter} were calculated through random sampling.
For each year $t$ in our data, denote the number of workers in the original sample by $N_t$.
First, for each year, we use a bootstrapping procedure to sample with replacement $N_t$ workers from the original data.
Next, we remove duplicated workers to create a sample without replacement.
These steps ensure that the percentage of workers that remain is randomly defined for each draw.
Then, we recompute the size of each firm, calculating the moments necessary for our calibration for the new sample.
The procedure was repeated 1,500 times for each year in the data.
After that, we calibrate the model and perform counterfactual analysis for each sample. %
Finally, we rank the results of the calibration and counterfactual analysis from the smallest to the largest.
The 2.5th and 97.5th percentile of the results simply represent the bounds of the 95\% confidence interval for the parameter or moment in question. %

The reason we perform bootstrapping on workers rather than on firms is because the latter would lead to an underestimation of the spread of within-firm variance.
We remove duplicated workers after bootstrapping because they would bias the within-firm variance estimator downward. %
Furthermore, we do not change worker-firm matching because randomly assigning workers to firms would undermine the sorting mechanism in the data.

For each sample, approximately two thirds of the workers and over 99\% of firms remain after each random sampling procedure.
Unweighted estimates are based on the firm sample, while weighted estimates are derived from the paired worker-firm sample.
To ensure variance estimators are unbiased, we use Bessel's correction, scaling the sum of squared mean deviations by $N - 1$ or $n - 1$, where $N$ denotes worker-firm pairs for overall variance, and $n$ represents firm size for within-firm variance.
We compute between-firm variance by subtracting within-firm variance from overall variance.
To derive the standard deviation of within-firm wage variance, we adjust for the uncertainty  in the estimation of the variance itself. Following, for example, \citet{ONeill2014-yx},
we use adjustment factor $\widehat{\mu}_4 / n - \widehat{\mu}_2 ^ 2 (n-3) / (n^2-n)$, where $\widehat{\mu}_k$ is the $k$'s sample moment for a given firm and $n$ is the firm size. We subtract this firm-specific adjustment factor from each firm's within-firm variance, and then calculate the variance of these adjusted within-firm variances. %

After the adjustment, our resampling procedure produces central estimates of the variance of within-firm variances that are very close to, but not identical to the adjusted moment obtained from the original sample (see Figure \ref{fig: datavsboot} below, which plots the central estimates derived from the data with the central estimates obtained from averaging over the 1500 bootstrap samples). Note that the raw estimates are comfortably contained within the confidence intervals produced by the resampling procedure.

\newpage

\section{Additional Figures}\label{sec: addfigures}

\begin{figure}[!h]
\begin{center}
\subfloat{\resizebox{\textwidth}{!}{\input{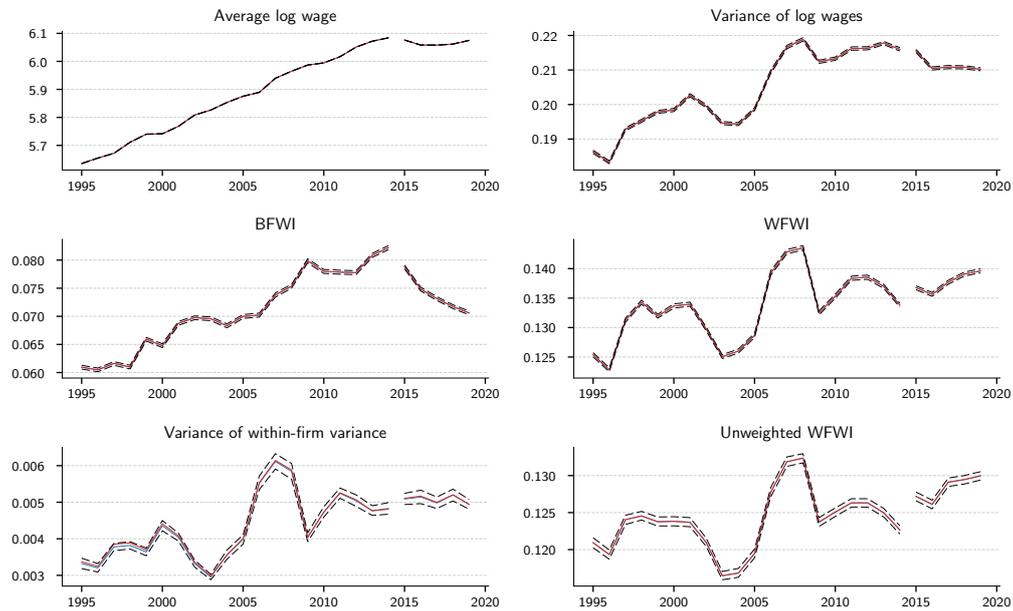}}}
\caption{Moments used in calibration}
\label{fig: datavsboot}
\end{center}

\footnotesize{\label{page: datavsboot} Notes: This plots the data moments from the original sample (red), the average over the 1500 bootstrap samples (blue) and the confidence interval (dashed black).  Top panel: (1) Mean log wage; (2) Variance of log wages. Middle panel: (1) The across firms, firm-size weighted variance of (within firm) mean log wages; (2) The across firms, firm-size weighted mean  of the variance of log wages. Bottom panel: (1) The across firm, firm-size weighted variance of the (within-firm) variance of log wages; (2) The across firms, size-unweighted mean  of the variance of log wages.}

\end{figure}

\vspace*{\fill} %

\clearpage
\begin{figure}[!h]
\begin{center}
\subfloat{\resizebox{\textwidth}{!}{\input{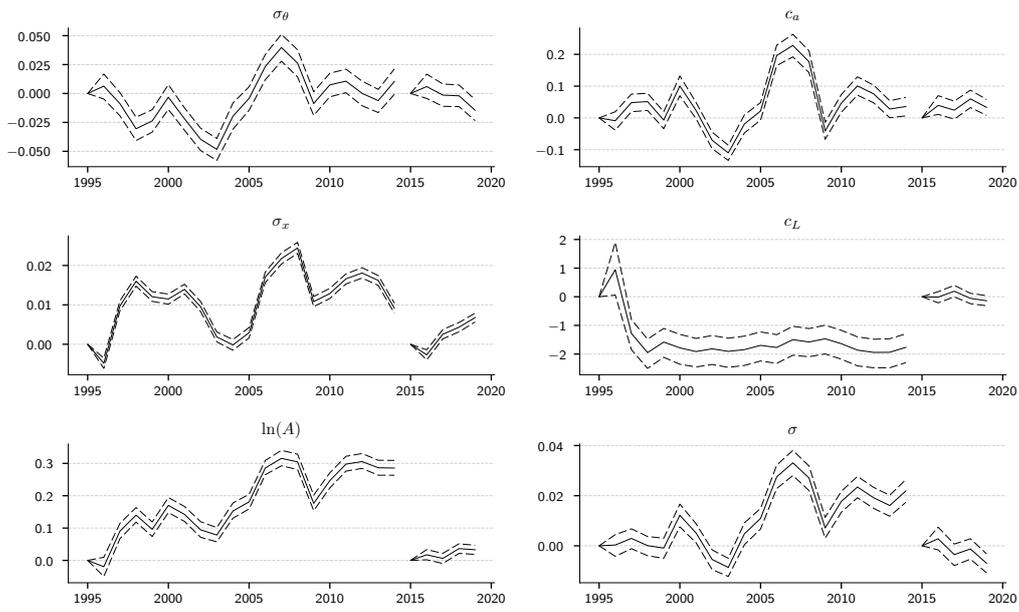}}}
\caption{Change in calibrated parameters}
\label{fig: diffstime}
\end{center}

\footnotesize{Notes: The difference in the calibrated parameters between the year in question and 1995. Top panel: (1) the variance of productivity $\sigma_\theta$, (2) cost of amenity provision $c_a$. Middle panel: (1) variance of skill $\sigma_x$, (2) span-of-control cost $c_l$. Bottom panel: (1) log of total factor productivity $\ln A$, (2) equilibrium supply of quality jobs $\sigma$. The dashed lines depict the 95\% confidence intervals.}

\end{figure}

\end{document}